\documentclass[prd,preprint,preprintnumbers,nofootinbib,eqsecnum,superscriptaddress]{revtex4}

 \usepackage[dvips,final]{graphicx}
  \usepackage{amssymb}
   \usepackage{amsmath}
    \usepackage{amsfonts}
     \usepackage{epsfig}
      \usepackage{bm}

\usepackage{mathpazo}

\usepackage[section]{placeins}

\input epsf.tex
\def\desepsf(#1 width #2){\epsfxsize=#2 \epsfbox{#1}}

\usepackage[normalem]{ulem}

\usepackage{multirow}
\usepackage{ctable}
\usepackage{booktabs}
\usepackage{array}
\usepackage{tabularx}
\usepackage{xcolor}
\usepackage{pstricks}
\definecolor{fred}{rgb}{0.90053, 0.00369, 0.00159}  

\newcommand{\be}{\begin{eqnarray}}
\newcommand{\ee}{\end{eqnarray}}

\begin{document}

\author{Rafa{\l} Maciu{\l}a}
\email{rafal.maciula@ifj.edu.pl}
\affiliation{Institute of Nuclear
Physics, Polish Academy of Sciences, ul. Radzikowskiego 152, PL-31-342 Krak{\'o}w, Poland}

\author{Antoni Szczurek\footnote{also at University of Rzesz\'ow, PL-35-959 Rzesz\'ow, Poland}}
\email{antoni.szczurek@ifj.edu.pl} 
\affiliation{Institute of Nuclear Physics, Polish Academy of Sciences, ul. Radzikowskiego 152, PL-31-342 Krak{\'o}w, Poland}



\title{Impact of the LHCb $\bm{p\!+\!\!^{4}\!H\!e}$ fixed-target  $\bm{D^0/\bar{D^0}}$ data
on \\the intrinsic $\bm{c \bar c}$ component in the nucleon}

\begin{abstract}
We discuss the impact of the new LHCb fixed-target $p\!+\!^{4}\!He$ data for
$D^0/{\bar D}^0$ production on the intrinsic $c \bar c$ component in
the nucleon wave function. Within the scenario presented here neither 
the traditional gluon-gluon fusion or quark-antiquark annihilation 
mechanisms calculated in the $k_T$-factorization
approach nor their counterparts from the collinear next-to-leading order collinear framework are sufficient to describe the transverse momentum and
rapidity distributions of $D^0/{\bar D}^0$ mesons. First the $c{\bar c}$-pair production within the standard frameworks is considered. Here a crucial role of the $c \to D$ hadronization effects at low energies and low transverse momenta is found and discussed, which was not analyzed in previous studies. A contribution related to intrinsic $c \bar c$ component in the nucleon wave function is included in addition.
Two models of the symmetric ($c(x) = \bar c(x)$) intrinsic charm (IC) component are
considered. The intrinsic charm $g^* c \to g c$ (or $g^* {\bar c} \to g {\bar c}$)
contribution needs to be regularized in order to obtain a suppression of the minijet $p_{T}$ spectrum present in the phenomenological minijet model, commonly used in Monte Carlo generators.
We show that in our model
the regularization parameter can be obtained from the fit to 
the LHCb fixed-target data under consideration here. 
We discuss uncertainties of our calculations (scale, charm quark mass, fragmentation function) as well as set limits on the IC probability.
According to our model the intrinsic charm probability $P_{IC} = 1.65\%$ allows to significantly improve description of the LHCb data
but the number is rather uncertain.
\end{abstract} 

\maketitle

\section{Introduction}

The proton is known to be a complicated object built of QCD degrees 
of freedom: quarks, antiquarks and gluons or even
their conglomerates. However, its wave function remains to large extent
unknown. The flavour structure of the nucleon is especially interesting.

An interesting $\bar d - \bar u$ asymmetry confirmed by the Gottfried
sum rule violation \cite{GSR_violation} and fixed target Drell-Yan
experiment \cite{Drell-Yan} is one example. This is usually
explained in terms of the pion (meson) cloud in the nucleon
(see e.g. Ref.~\cite{HSS}). The pion cloud probability is of the order
of 30\%. It contributes to nonperturbative light quark/antiquark
distributions in the nucleon.
Another example is the strangeness in the nucleon. Here 
the kaon cloud leads to a nonperturbative contribution of $s / \bar s$
quarks/antiquarks.
A few percent effect of the nonperturbative $s \bar s$ component 
was predicted \cite{HSS}.

The situation for the charm-anticharm content of the nucleon
is less understood.
Long ago Brodsky et al. \cite{BHPS1980} proposed a simple model of 
the large-$x$ intrinsic charm due to $u u d c \bar c$ Fock component 
of the wave function. As a consequence of heavy mass of the $c$-quark 
or $\bar c$-antiquark their longitudinal momentum fractions are of the order
of $x \sim 0.3-0.5$. In its simplest form the model leads to
$c(x) = {\bar c}(x)$ (symmetric IC). Another possible mechanism is
nonperturbative gluon splitting into $c \bar c$.
Such a mechanism was discussed e.g. in 
Refs.~\cite{Edin:1998dz,Maciula:2020dxv}.
Also here $c(x) = {\bar c}(x)$, but typical longitudinal momentum
fractions of $c$-quark or ${\bar c}$-antiquark are much smaller than in 
the Brodsky-Hoyer-Peterson-Sakai (BHPS) model.
We shall call the second model as the SEA-like.
It is very difficult to calculate in a realistic way
the integral $\int c(x) dx = \int {\bar c}(x) dx$ (probability of
such a component in the wave function) as it goes beyond the scope
of perturbative methods.

In Ref.~\cite{Melnitchouk:1997ig} the $c(x)$ and $\bar c(x)$ instrinsic 
charm/anticharm distributions are calculated postulating heavy charmed 
meson - heavy charmed baryon component in the nucleon wave function. 
In this approach typically $c(x)$ is somewhat different than 
${\bar c}(x)$ (asymmetric IC) but the most probable range of $x$ is 
similar as for the BHPS model.
Therefore our analysis of the BHPS model is approximately valid
also for the meson cloud approach.

The intrinsic charm component was discussed recently in Refs.~\cite{Bednyakov:2013zta,Bednyakov:2017vck,Brodsky:2020zdq}
in the context of the associated production of prompt photons and charm-quark jets in $pp$-collisions at LHC energies.
In Refs.~\cite{Beauchemin:2014rya,Lipatov:2016feu,Lipatov:2018oxm} an associated production of $Z^0$ and heavy flavoured (HF) jets was studied
in the context of identifying intrinsic charm in the nucleon. Recently, in Ref.~\cite{Lipatov:2018oxm} the authors found that the forthcoming ATLAS and CMS measurements of $Z^0+\mathrm{HF}$ production at $\sqrt{s}=13$ TeV can be very important in the context of searching for the IC contribution in the proton. They suggested measuring a new observable, defined as the double ratio of cross sections $\left(\frac{\sigma^{forward}(Z+c)}{\sigma^{central}(Z+c)}\right)/\left(\frac{\sigma^{forward}(Z+b)}{\sigma^{central}(Z+b)}\right)$ (where central and forward means $|y_{Z}|<1.5$ and $1.5 < |y_{Z}| < 2.5$, respectively). This was found to be extremely sensitive to the IC signal, which would be potentially visible at large transverse momenta of $Z^{0}$-boson and/or heavy flavour jet. 
At low energy discussed in the present paper the production of $Z^0$ is not possible.
On the other hand registration of the extra photon is not possible due to very low
luminosity in the fixed target experiment.

In the present study we are interested rather in production
of $c$-quark and/or ${\bar c}$-antiquark and associated $D$ mesons 
in the forward direction.
The interaction of gluon (small-$x$) from the projectile 
with $c$ or $\bar c$ (large-$x$) in the target ($^{4}\!He$) is 
the underlying partonic mechanism \cite{Maciula:2020dxv}. 
The $^{4}\!He$ is rather light nucleus 
and the associated nuclear effects are of the order of a few percent
only and can be safely neglected. We shall discuss this issue shortly
in the main text.
Focusing on rather small charm transverse momenta, the $cg^* \to cg$ (or $\bar{c}g^* \to \bar{c}g$) subprocess needs to be regularized in order to obtain a phenomenologically motivated suppression of the minijet $p_{T}$ spectrum.
Here we follow the minijet model originally proposed in Ref.~\cite{Sjostrand:1987su} and further adopted in modern Monte Carlo generators (see e.g. \textsc{Pythia} \cite{Sjostrand:2014zea}).
The parameter governing the regularization must be adjusted to
experimental data. It will be discussed how to do it in our case.

 
In parallel we have made a similar analysis of the IC signal for high-energy prompt
neutrino production in the Earth's atmosphere coming from the decay 
of charmed meson \cite{Goncalves:2021yvw}.
The two analyses (here and for atmospheric neutrino) are complementary. Combining conclusions from
both of them may be therefore very useful.

\section{Details of the model calculations}

\subsection{The leading perturbative mechanism}

\begin{figure}[!h]
\centering
\begin{minipage}{0.3\textwidth}
  \centerline{\includegraphics[width=1.0\textwidth]{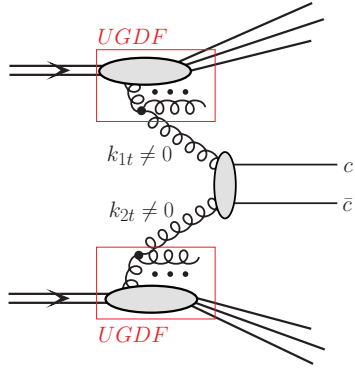}}
\end{minipage}
  \caption{
\small A diagramatic representation of the leading perturbative mechanism of charm production in the $k_{T}$-factorization approach driven by the fusion of two off-shell gluons.
}
\label{fig:diagramLO}
\end{figure}

We remind here very briefly the theoretical formalism for the calculation of the $c\bar{c}$-pair production in the $k_{T}$-factorization approach \cite{kTfactorization}, as adopted and discussed by one of us in the context of the LHCb fixed-target charm data in Ref.~\cite{Maciula:2020cfy}. In this framework the transverse momenta $k_{t}$'s (virtualities) of both partons entering the hard process are taken into account, both in the matrix elements and in the parton distribution functions. Emission of the initial state partons is encoded in the transverse-momentum-dependent (unintegrated) PDFs (uPDFs). In the case of charm flavour production the parton-level cross section is usually calculated via the $2\to 2$ leading-order $g^*g^* \to c\bar c$ fusion mechanism with off-shell initial state gluons that is the dominant process at high energies (see Fig.~\ref{fig:diagramLO}). Even at lower energies as long as small transverse momenta and not extremely backward/forward rapidities are considered the $q^*\bar q^* \to c\bar c $ mechanism remains subleading. Then the hadron-level differential cross section for the $c \bar c$-pair production, formally at leading-order, reads:
\begin{eqnarray}\label{LO_kt-factorization} 
\frac{d \sigma(p p \to c \bar c \, X)}{d y_1 d y_2 d^2p_{1,t} d^2p_{2,t}} &=&
\int \frac{d^2 k_{1,t}}{\pi} \frac{d^2 k_{2,t}}{\pi}
\frac{1}{16 \pi^2 (x_1 x_2 s)^2} \; \overline{ | {\cal M}^{\mathrm{off-shell}}_{g^* g^* \to c \bar c} |^2}
 \\  
&& \times  \; \delta^{2} \left( \vec{k}_{1,t} + \vec{k}_{2,t} 
                 - \vec{p}_{1,t} - \vec{p}_{2,t} \right) \;
{\cal F}_g(x_1,k_{1,t}^2,\mu_{F}^2) \; {\cal F}_g(x_2,k_{2,t}^2,\mu_{F}^2) \; \nonumber ,   
\end{eqnarray}
where ${\cal F}_g(x_1,k_{1,t}^2,\mu_{F}^2)$ and ${\cal F}_g(x_2,k_{2,t}^2,\mu_{F}^2)$
are the gluon uPDFs for both colliding hadrons and ${\cal M}^{\mathrm{off-shell}}_{g^* g^* \to c \bar c}$ is the off-shell matrix element for the hard subprocess.
The gluon uPDF depends on gluon longitudinal momentum fraction $x$, transverse momentum
squared $k_t^2$ of the gluons entering the hard process, and in general also on a (factorization) scale of the hard process $\mu_{F}^2$.
The extra integration is over transverse momenta of the initial
partons. Here, one keeps exact kinematics from the very beginning and additional hard dynamics coming from transverse momenta of incident partons. Explicit treatment of the transverse momenta makes the approach very efficient in studies of correlation observables. The two-dimensional Dirac delta function assures momentum conservation. The gluon uPDFs must be evaluated at longitudinal momentum fractions 
$x_1 = \frac{m_{1,t}}{\sqrt{s}}\exp( y_1) + \frac{m_{2,t}}{\sqrt{s}}\exp( y_2)$, and $x_2 = \frac{m_{1,t}}{\sqrt{s}}\exp(-y_1) + \frac{m_{2,t}}{\sqrt{s}}\exp(-y_2)$, where $m_{i,t} = \sqrt{p_{i,t}^2 + m_c^2}$ is the quark/antiquark transverse mass.  

As we have carefully discussed in Ref.~\cite{Maciula:2019izq}, there is a direct relation between a resummation present in uPDFs in the transverse momentum dependent factorization and a parton shower in the collinear framework. Such observations have been made also earlier, for example in Refs.~\cite{Bury:2016cue,Bury:2017jxo}. The commonly accepted statement is that actually in the $k_{T}$-factorization approach already at leading-order some part of radiative higher-order corrections can be effectively included via uPDFs, without any additional showering procedure. However, it is true only for those uPDF models in which extra emissions of soft and even hard partons are encoded, including $k_{t}^{2} > \mu_{F}^{2}$ configurations. In most uPDF cases the off-shell gluon can be produced either from gluon or quark, therefore, in the $k_{T}$-factorization all channels driven by $gg, q\bar q$ and even by $qg$ initial states are open already at leading-order (in contrast to the collinear factorization). Then, when calculating the charm production cross section via the $g^* g^* \to c \bar c$ mechanism one could expect to effectively include contributions related to an additional one or two (or even more) extra partonic emissions which in some sense plays a role of the initial state parton shower.

The kinematical configuration of the fixed-target LHCb experiment corresponds to the region where in principle the Catani-Ciafaloni-Fiorani-Marchesini (CCFM) \cite{CCFM} evolution equation is legitimate for any pQCD theoretical calculations and could, in principle, be used to describe the dynamics behind the mechanisms of \textit{e.g.} open charm meson production. Within the CCFM approach the parton transverse momentum is allowed to be larger than the scale $\mu^2$. This useful feature translates into the easy and effective taking into account of higher-order radiative corrections, that correspond to the initial-state real gluon emissions which are resummed into the uPDFs. In the numerical calculations below we follow the conclusions from Ref.~\cite{Maciula:2020cfy} and apply the gluon uPDFs obtained from the CCFM evolution equation. There a different models of the CCFM unintegrated gluon densities have been tested and found to lead to a reasonable description of the LHCb fixed-target charm data.
The best theory to data relation has been obtained in the case of the most up-to-date JH-2013-set1 and JH-2013-set2 unintegrated gluon densities \cite{Hautmann:2013tba} that are determined from high-precision DIS measurements.

As a default set in the numerical calculations we take the renormalization scale
$\mu^2 = \mu_{R}^{2} = \sum_{i=1}^{n} \frac{m^{2}_{it}}{n}$ (averaged transverse mass of the given final state) and the charm quark mass $m_{c}=1.5$ GeV. The strong-coupling constant $\alpha_{s}(\mu_{R}^{2})$ at next-to-next-to-leading-order is taken from the CT14nnloIC PDF routines\footnote{In the numerical calculations in Ref.~\cite{Maciula:2020cfy} the strong-coupling at leading-order was taken. Here, in order to consistently compare the standard mechanism and the intrinsic charm component we keep in both cases the $\alpha_{s}(\mu_{R}^{2})$ values encoded in the CT14nnloIC PDF.}\cite{Hou:2017khm}. The CCFM uPDFs here are taken at rather atypical value of the factorization scale $\mu_{F}^2 = M_{c\bar c}^2 + P_{T}^{2}  $, where $M_{c\bar c}$ and $P_{T}$ are the $c\bar c$-invariant mass (or energy of the scattering subprocess) and the transverse momentum of $c\bar c$-pair (or the incoming off-shell gluon pair). This unusual definition has to be applied as a consequence of the CCFM evolution algorithm \cite{Hautmann:2013tba}. For completeness of the present study, in some places we also apply the Martin-Ryskin-Watt (MRW)
model for unintegrated gluon densities \cite{Watt:2003mx}.

\subsection{The intrinsic charm contribution}

\begin{figure}[!h]
\centering
\begin{minipage}{0.3\textwidth}
  \centerline{\includegraphics[width=1.0\textwidth]{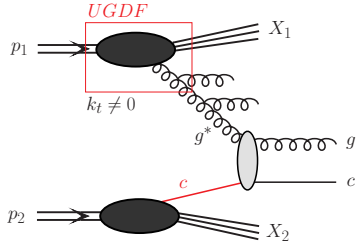}}
\end{minipage}
  \caption{
\small A diagrammatic representation of the intrinsic charm mechanism of charm production within the hybrid model with the off-shell gluon and the on-shell charm quark in the initial state.
}
\label{fig:diagramIC}
\end{figure}

The intrinsic charm contribution to charm production cross section (see Fig.~\ref{fig:diagramIC}) is obtained within the hybrid theoretical model discussed by us in detail in Ref.~\cite{Maciula:2020dxv}. The LHCb fixed-target configuration allows to explore the charm cross section in the backward rapidity direction where an asymmetric kinematical regime can be explored. Thus in the basic $gc \to gc$ reaction the gluon PDF and the intrinsic charm PDF are simultaneously probed at different longitudinal momentum fractions - rather intermediate for the gluon and large for the charm quark.

Within the asymmetric kinematic situation $x_1 \ll x_2$ the cross section for the processes under consideration can be calculated in the so-called hybrid factorization model motivated by the work in Ref.~\cite{Deak:2009xt}. In this framework the small- or intermediate-$x$ gluon is taken to be off mass shell and 
the differential cross section e.g. for $pp \to g c X$ via $g^* c \to g c$ mechanism reads:
\begin{eqnarray}
d \sigma_{pp \to gc X} = \int d^ 2 k_{t} \int \frac{dx_1}{x_1} \int dx_2 \;
{\cal F}_{g^{*}}(x_1, k_{t}^{2}, \mu^2) \; c(x_2, \mu^2) \; d\hat{\sigma}_{g^{*}c \to gc} \; ,
\end{eqnarray}
where ${\cal F}_{g^{*}}(x_1, k_{t}^{2}, \mu^2)$ is the unintegrated gluon distribution in one proton and $c(x_2, \mu^2)$ a collinear PDF in the second one. The $d\hat{\sigma}_{g^{*}c \to gc}$ is the hard partonic cross section obtained from a gauge invariant tree-level off-shell amplitude. A derivation of the hybrid factorization from the dilute limit of the Color Glass Condensate approach can be found in Ref.~\cite{Dumitru:2005gt} (see also Ref.~\cite{Kotko:2015ura}). The relevant cross sections are calculated with the help of the \textsc{KaTie} Monte Carlo generator \cite{vanHameren:2016kkz}. There the initial state quarks (including heavy quarks) can be treated as a massless partons only.  

Working with minijets (jets with transverse momentum of the order of a few GeV) requires a phenomenologically motivated regularization of the cross sections. Here we follow the minijet model \cite{Sjostrand:1987su} adopted e.g. in \textsc{Pythia} Monte Carlo generator, where a special suppression factor is introduced at the cross section level \cite{Sjostrand:2014zea}:
\begin{equation}
F(p_t) = \frac{p_t^2}{ p_{T0}^2 + p_t^2 } \; 
\label{Phytia_formfactor}
\end{equation}
for each of the outgoing massless partons with transverse momentum $p_t$, where $p_{T0}$ is a free parameter of the form factor
that also enters as an argument of the strong coupling constant $\alpha_{S}(p_{T0}^2+\mu_{R}^{2})$.

This suppression factor was originally proposed to remove singularity of minijet cross sections in the collinear approach at leading-order. In the hybrid model (or in the $k_{T}$-factorization) the leading-order cross sections are finite as long as $k_{T}> 0$, where $k_{T}$ is the transverse momentum of the incident off-shell parton. Within this approach, a treatment of the small-$k_{T}$ region in the construction of a given unintegrated parton density is crucial. Different models of uPDFs may lead to different behaviour of the cross section at small minijet transverse momenta but in any case the cross sections should be finite. However, as it was shown in Ref.~\cite{Kotko:2016lej}, the internal $k_{T}$ cannot give a minijet suppression consistent with the minijet model and the related regularization seems to be necessary even in this framework.

\begin{figure}[!h]
\begin{minipage}{0.33\textwidth}
  \centerline{\includegraphics[width=1.0\textwidth]{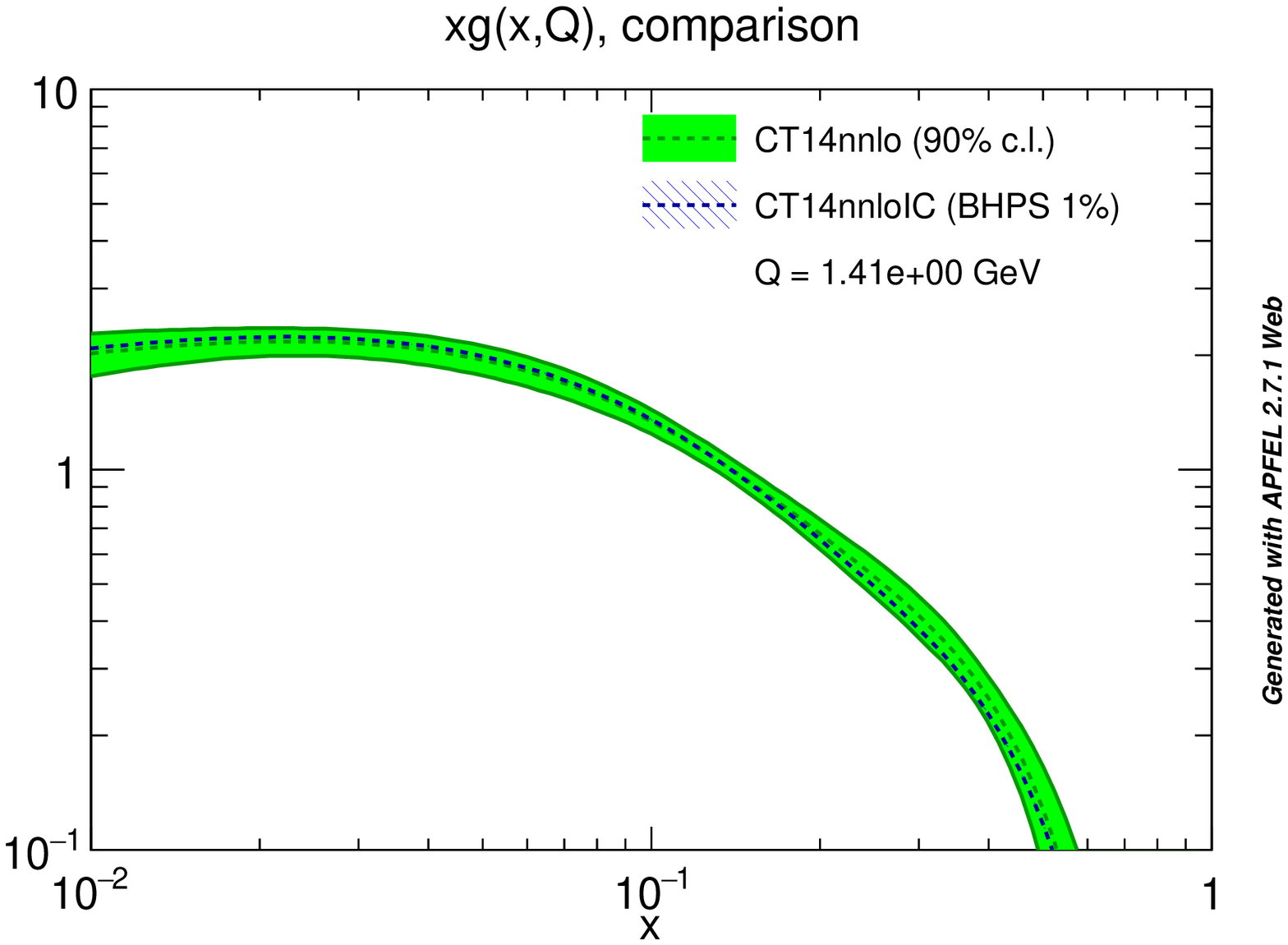}}
\end{minipage}
\begin{minipage}{0.33\textwidth}
  \centerline{\includegraphics[width=1.0\textwidth]{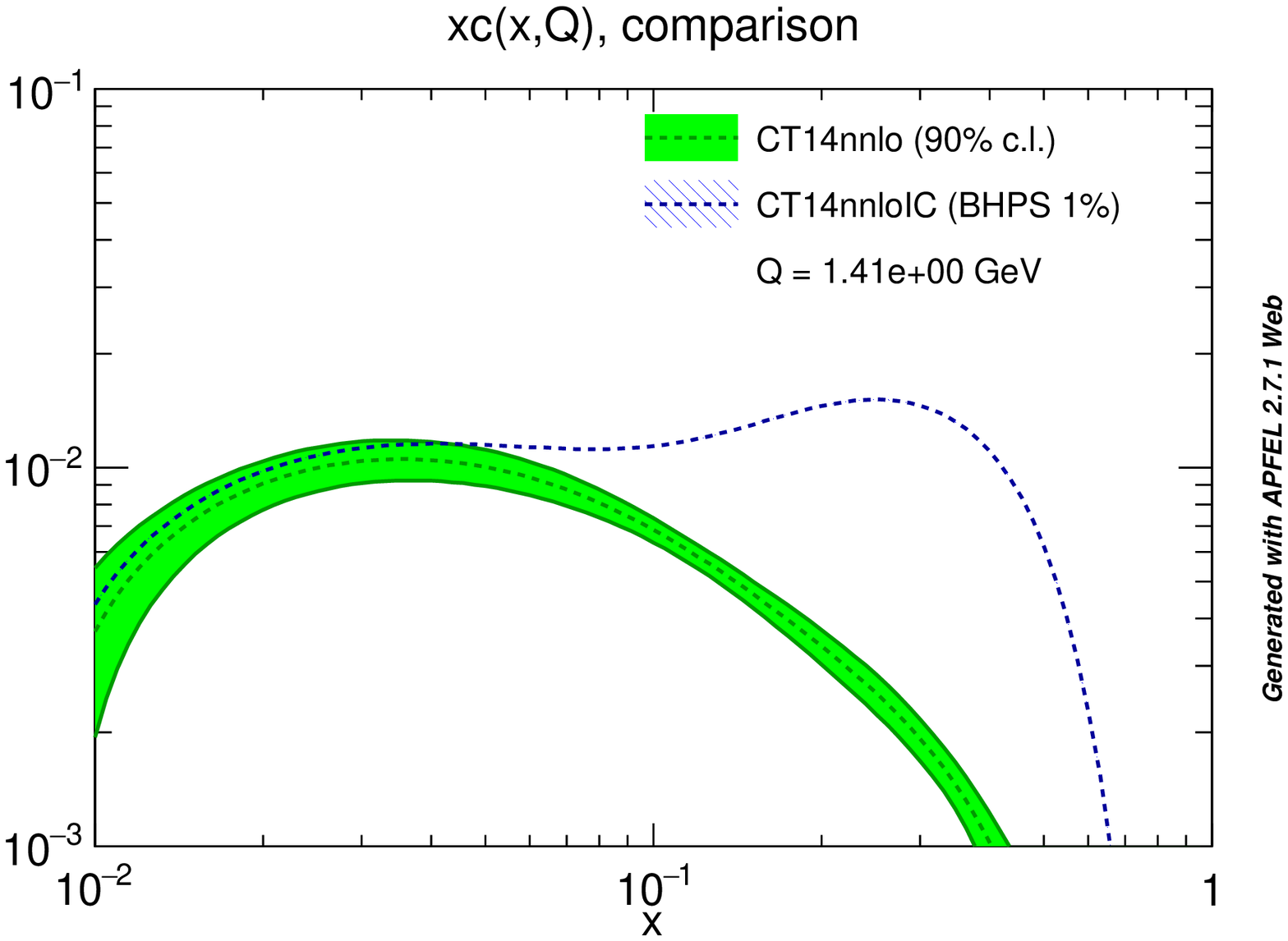}}
\end{minipage}\\
\begin{minipage}{0.33\textwidth}
  \centerline{\includegraphics[width=1.0\textwidth]{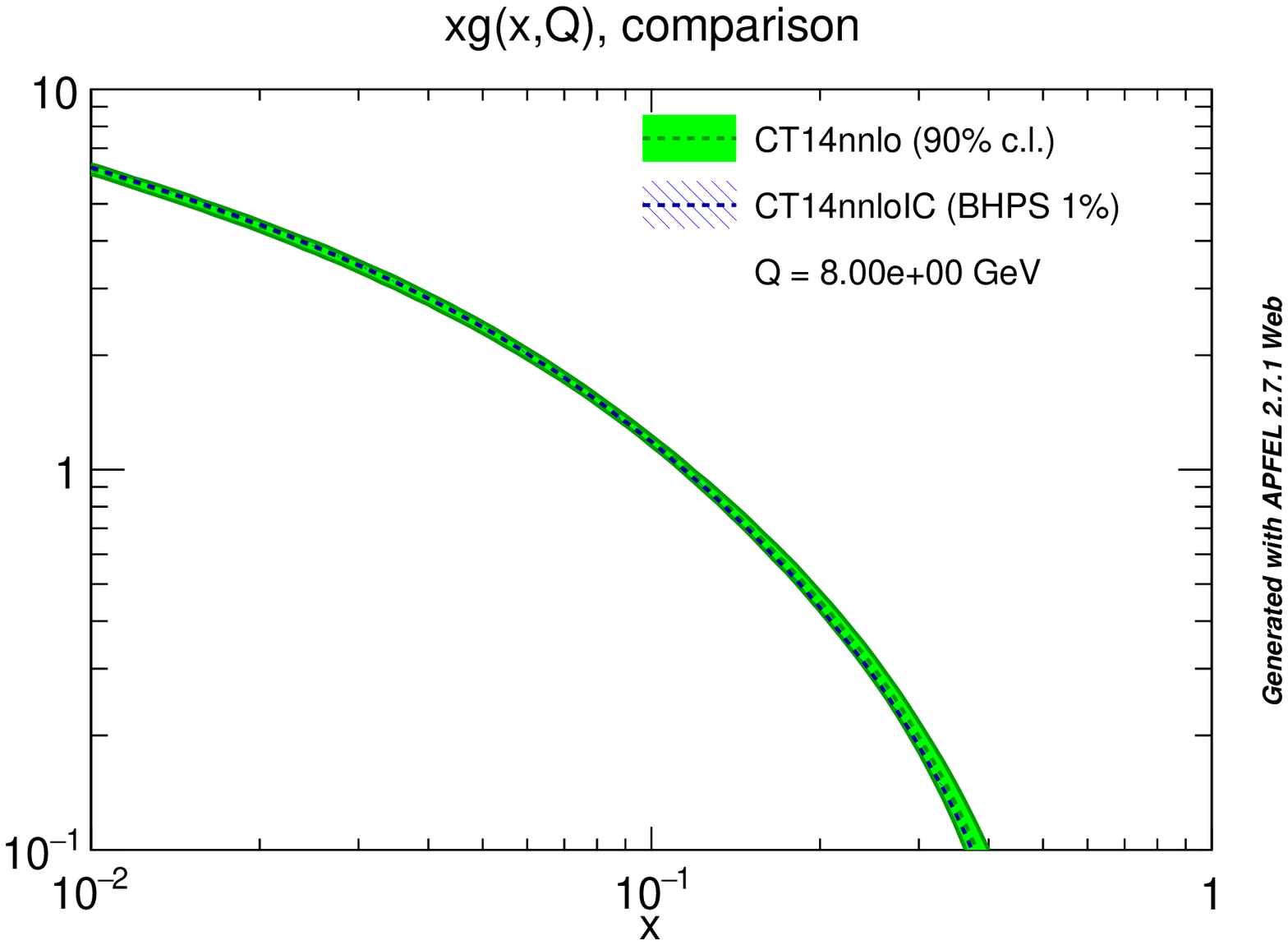}}
\end{minipage}
\begin{minipage}{0.33\textwidth}
  \centerline{\includegraphics[width=1.0\textwidth]{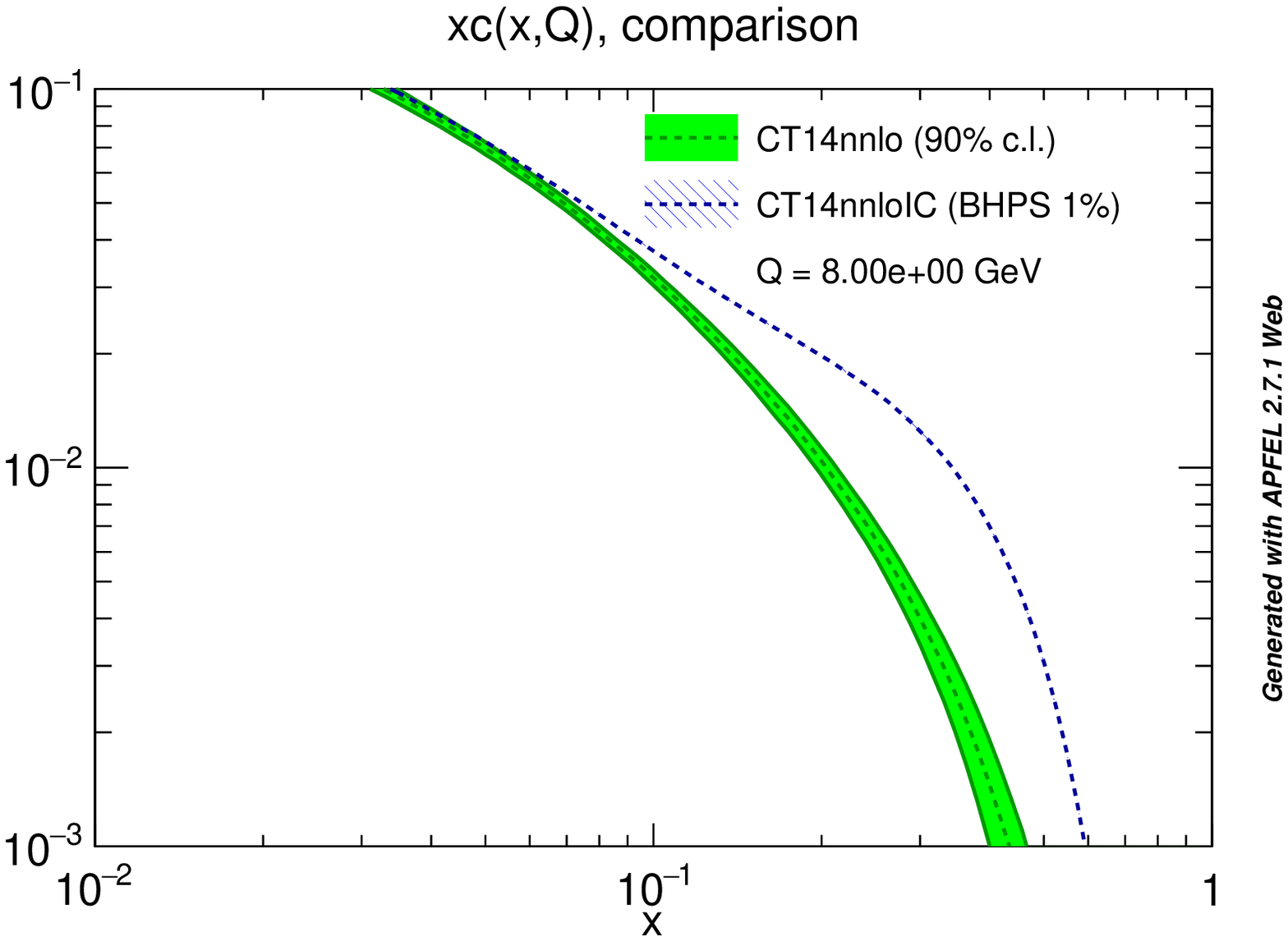}}
\end{minipage}
  \caption{
\small The gluon (left panels) and charm (right panels) collinear distributions in a proton from the CT14nnlo PDFs (shaded bands) without intrinsic charm and from the CT14nnloIC PDFs (dashed lines) with intrinsic charm. Upper and lower panels correspond to different scales. Here the BHPS 1\% model for intrinsic charm is taken. 
}
\label{fig:1}
\end{figure}

In the numerical calculations below, the intrinsic charm PDFs are taken at the initial scale $m_{c} = 1.3$ GeV, so the perturbative charm contribution is intentionally not taken into account. In the numerical calculations below we apply different grids of the intrinsic charm distributions from the CT14nnloIC PDF \cite{Hou:2017khm} that correspond to the BHPS model \cite{BHPS1980} as well as to the so-called SEA-like models.    
In Fig.~\ref{fig:1} we compare the collinear gluon (left panels) ant the charm quark (right panels) PDFs obtained with (CT14nnloIC) and without (CT14nnlo) the intrinsic charm concept at different factorization scales (upper and lower panels). Here, the intrinsic charm grid that correspond to the BHPS model with $1\%$ probability for intrinsic charm was used. In this case the intrinsic charm contribution leads to a significant enhancement of the charm PDF in the region of longitudinal momentum fraction $x > 0.1$.   
    
\subsection{The charm quark to meson transition}

The transition of charm quarks to open charm mesons is done in the framework of the independent parton fragmentation picture (see \textit{e.g.} Refs.~\cite{Maciula:2015kea,Maciula:2019iak}).
In Ref.~\cite{Maciula:2020cfy} we followed the standard prescription, where the inclusive distributions of open charm meson were obtained through a convolution of inclusive distributions of charm quarks/antiquarks and $c \to D$ fragmentation functions:
\begin{equation}
\frac{d \sigma(pp \rightarrow D X)}{d y_D d^2 p_{t,D}} \approx
\int_0^1 \frac{dz}{z^2} D_{c \to D}(z)
\frac{d \sigma(pp \rightarrow c X)}{d y_c d^2 p_{t,c}}
\Bigg\vert_{y_c = y_D \atop p_{t,c} = p_{t,D}/z} \;,
\label{Q_to_h}
\end{equation}
where $p_{t,c} = \frac{p_{t,D}}{z}$ and $z \in (0,1)$.
There the typical approximation was done that $y_c$ is unchanged in the fragmentation process, i.e. $y_D=y_c$. This commonly accepted and frequently used method was originally proposed for light partons. It can be safely used only when both, mass of the parton and mass of the hadron can be neglected \cite{Maciula:2015kea}.

In principle, this approximation may not be valid for the case of heavy and even light parton fragmentation to heavy object, especially, at lower energies or/and considering regions of small meson transverse momenta. It is obvious that working with massive particles this model may break down when approaching $p_T \sim m_D$ region. In this regime one could expect a violation of ”energy conservation”  and events with hadrons that have larger energies than the energy of the parent parton can frequently appear. In some corners of the phase space the $E_D < E_c$ relation may be broken very strongly. As long as one is considering c.m.s. midrapidities and/or large c.m.s. collision energies this mass effect shall be rather negligible, especially, when a low transverse momentum cut is applied. However, the situation may dramatically change when going to lower energies and discussing backward or forward production. This might be exactly the case of the LHCb fixed-target charm data.

Therefore, here we also follow a different idea and assume that the $D$-meson is emitted in the direction of parent $c$-quark/antiquark, i.e. $\eta_D=\eta_c$ (the same pseudorapidities or polar angles). Within this approach still different options for $z$-scaling come into game, including e.g. the three-momentum ($p_{c} = \frac{p_{D}}{z}$) and the light-cone momentum ($p^{+}_{c} = \frac{p^{+}_{D}}{z}$ where $p^{+} = E + p$) scaling \cite{Maciula:2019iak}.

In numerical calculations we take the Peterson fragmentation function \cite{Peterson:1982ak} with $\varepsilon = 0.05$, often used in the context of hadronization of heavy flavours. Then, the hadronic cross section is normalized by the relevant charm fragmentation fractions for a given type of $D$ meson \cite{Lisovyi:2015uqa}.  
In the numerical calculations below for $c \to D^{0}$ meson transition we take the fragmentation probability $\mathrm{P}_{c \to D} = 61\%$.

\section{Numerical results}

\begin{figure}[!h]
\begin{minipage}{0.47\textwidth}
  \centerline{\includegraphics[width=1.0\textwidth]{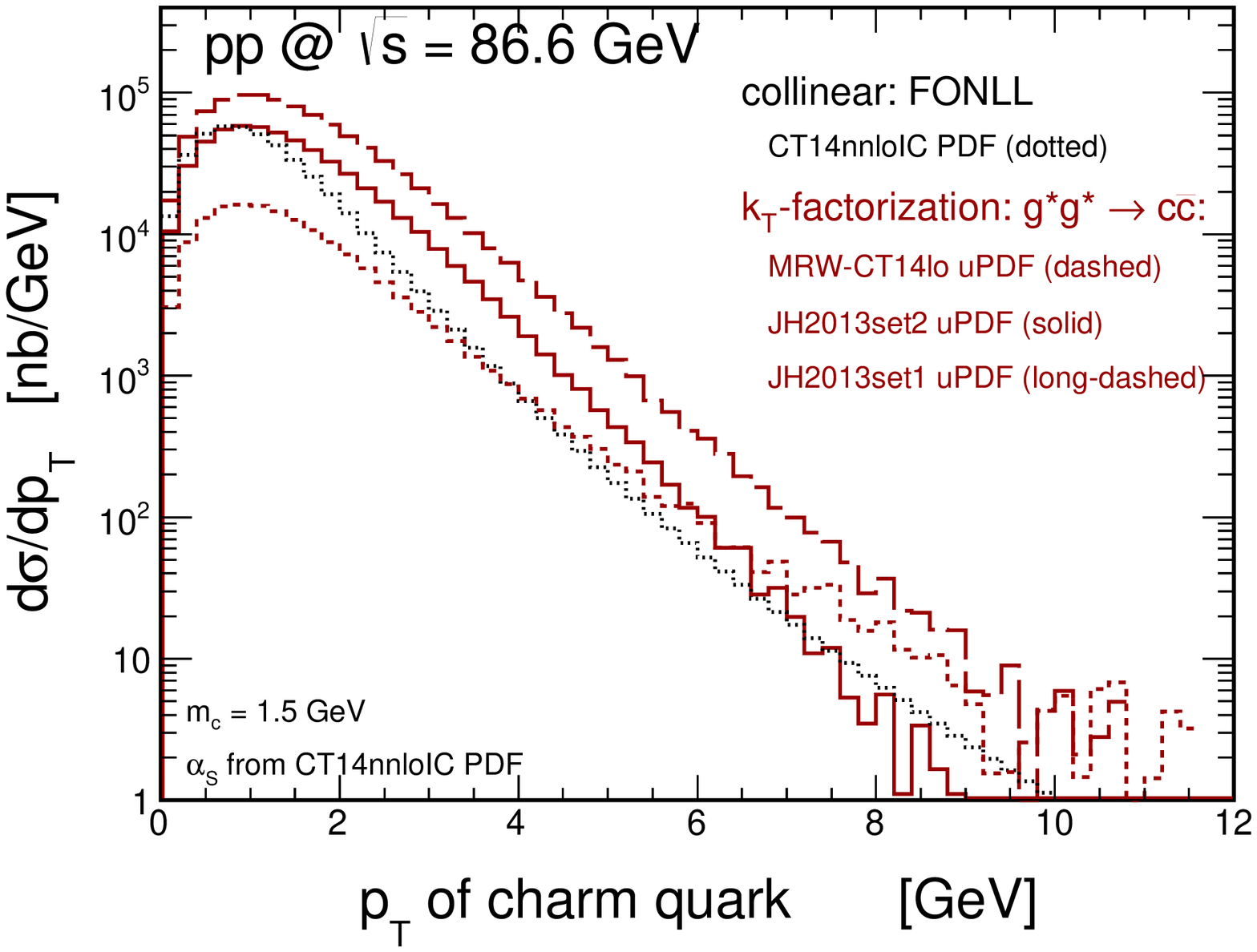}}
\end{minipage}
\begin{minipage}{0.47\textwidth}
  \centerline{\includegraphics[width=1.0\textwidth]{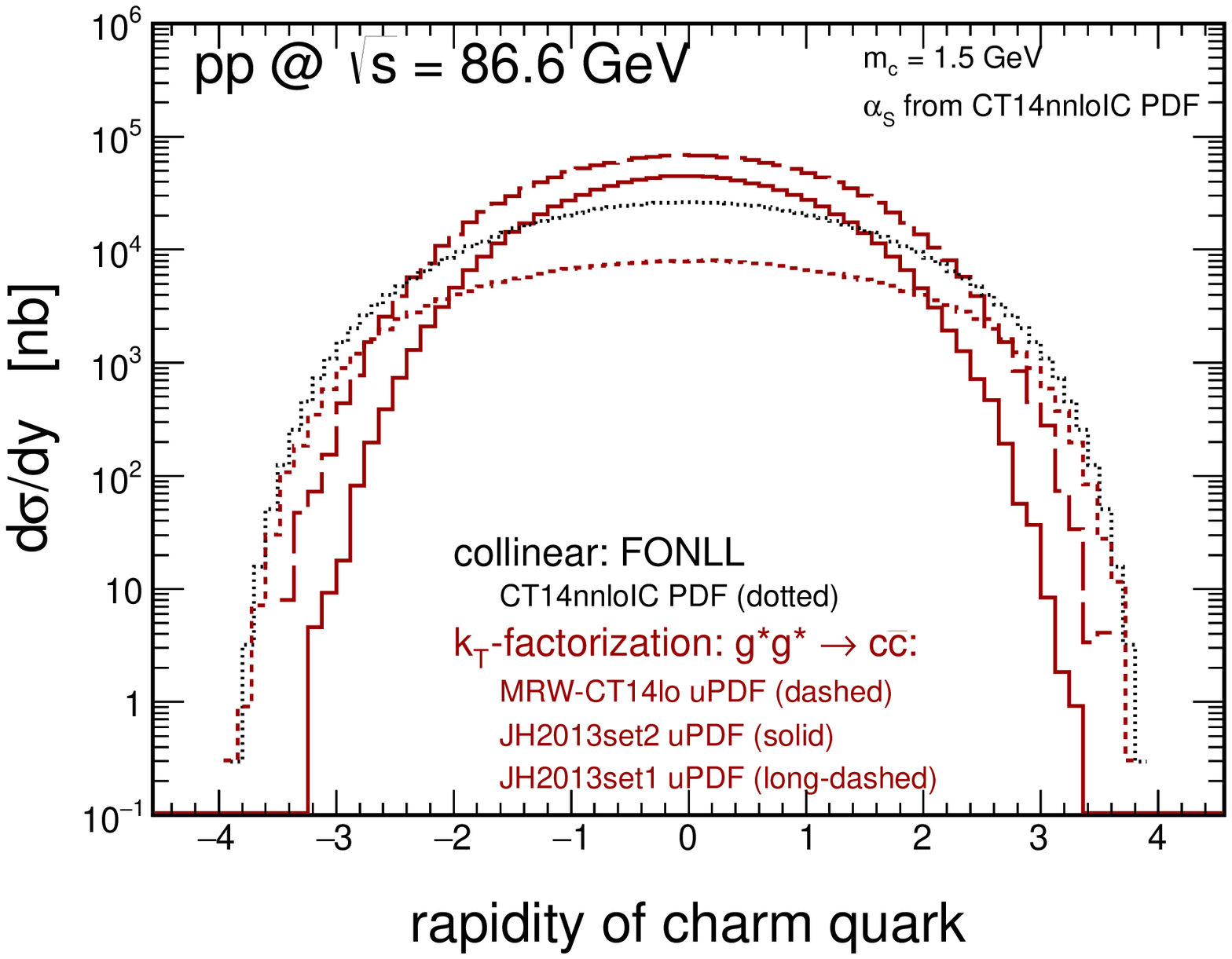}}
\end{minipage}\\
\begin{minipage}{0.47\textwidth}
  \centerline{\includegraphics[width=1.0\textwidth]{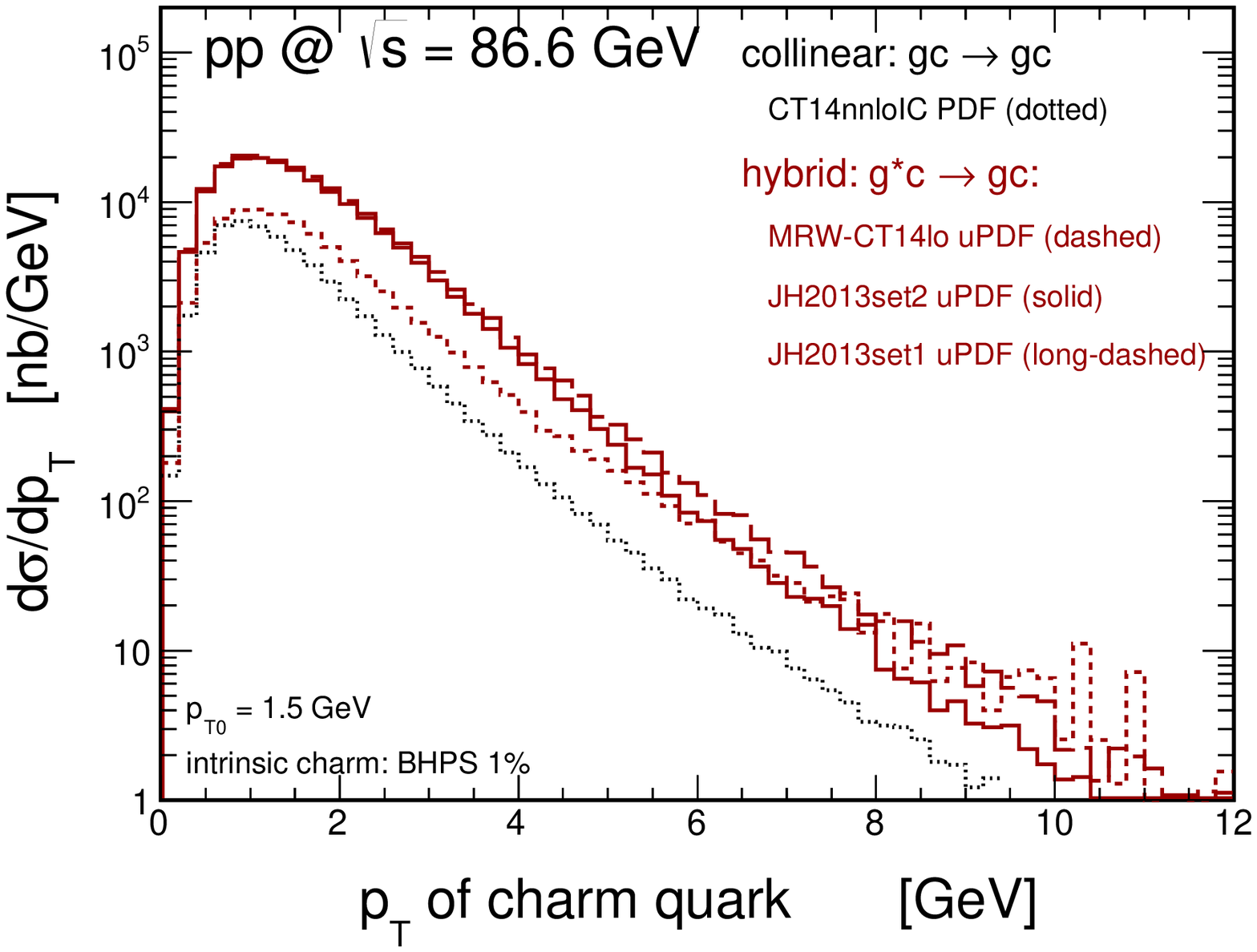}}
\end{minipage}
\begin{minipage}{0.47\textwidth}
  \centerline{\includegraphics[width=1.0\textwidth]{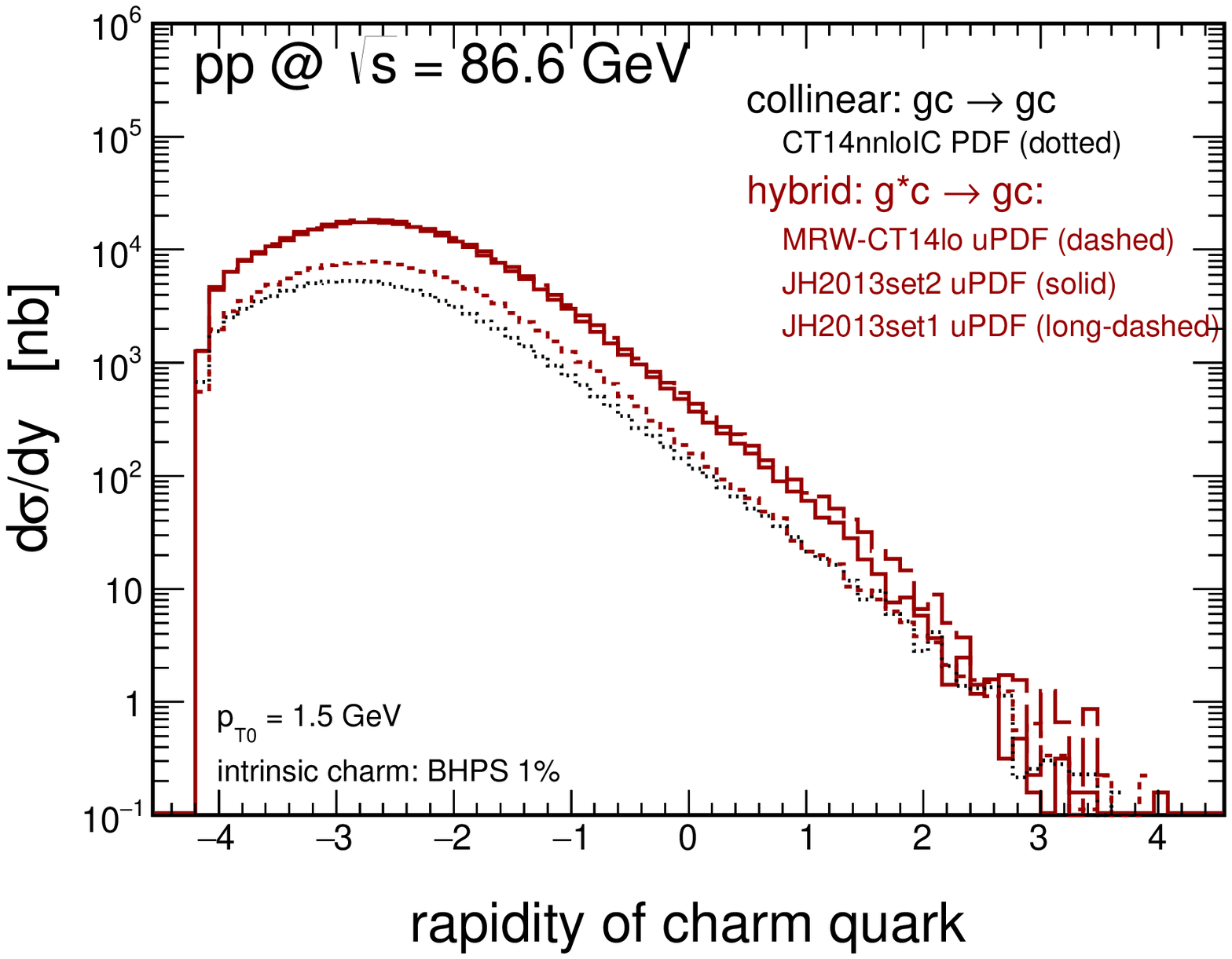}}
\end{minipage}
  \caption{
\small The transverse momentum (left) and rapidity (right) distributions of charm quark
for $pp$-collisions at $\sqrt{s} = 86.6$ GeV. Results for the $gg\to c\bar{c}$ (top panels) and the $gc\to gc$ (bottom panels) mechanisms are shown. 
The calculations for the former are done within the hybrid model as well as within the leading-order collinear approach. The latter results are obtained within the $k_{T}$-factorization and the FONLL frameworks. Here the BHPS 1\% model of intrinsic charm was used. Details are specified in the figure.  
}
\label{fig:2}
\end{figure}

Let us start presentation of our numerical results with the differential cross sections for charm quark production
in $pp$-scattering at $\sqrt{s} = 86.6$ GeV. In Fig.~\ref{fig:2} we show the transverse momentum (left panels) and rapidity (right panels) distributions of charm quark
for the standard $gg \to c\bar c$ partonic mechanism (top panels) as well as for the $gc \to gc$ mechanism with intrinsic charm in the initial state (bottom panels). Considering the standard charm production mechanism the results are obtained within the $k_{T}$-factorization approach for different models of gluon uPDFs, namely the JH2013set1 (long-dashed histograms), the JH2013set2 (solid histograms), and the MRW-CT14lo (dashed histograms). These results are compared in addition to the next-to-leading order collinear predictions obtained according to the FONLL \cite{Cacciari:1998it} framework (dotted histograms). Numerical results for the mechanism driven by the intrinsic charm are obtained within the hybrid approach with off-shell gluon and with collinear intrinsic charm distribution taken from the CT14nnloIC PDFs. Here we use the PDF set that correspond to the BHPS model with 1\% probability for finding intrinsic charm content in a proton. Here, the hybrid model results are compared with the leading-order collinear predictions. We observe that the different models of the gluon uPDF lead to quite different charm quark distributions. The statement is valid for both considered partonic mechanisms. Clearly, the unintegrated gluon densities are a source of large uncertainties of the predictions (see also discussion in Ref.~\cite{Maciula:2020dxv}). 
Therefore their application in the probed kinematics needs an experimental verification. In the case of the standard charm production mechanism it has been already done very recently in Ref.~\cite{Maciula:2020cfy}.

A large difference between transverse momentum distributions obtained with the MRW-CT14lo and the CCFM uPDFs, at small and intermediate transverse momenta of charm quark, comes from a quite different treatment of the small transverse momenta ($k_{t}$'s) of the incoming gluons within the two uPDF models. The resulting charm quark $p_{T}$ spectra are very sensitive to this region of the gluon $k_{t}$. The most up-to-date CCFM gluon uPDFs used here are determined from high-precision DIS measurements and put therefore a direct constraint on the small transverse momenta of gluons. In the MRW approach
a phenomenological model has to be applied for the treatment of the gluon uPDF in the $k_{t} \lesssim$ 2.0 GeV region, what makes the model less certain. In fact, at small scales (typical for the reactions considered here) even the collinear parton distributions are quite uncertain.  

\begin{figure}[!h]
\begin{minipage}{0.33\textwidth}
  \centerline{\includegraphics[width=1.0\textwidth]{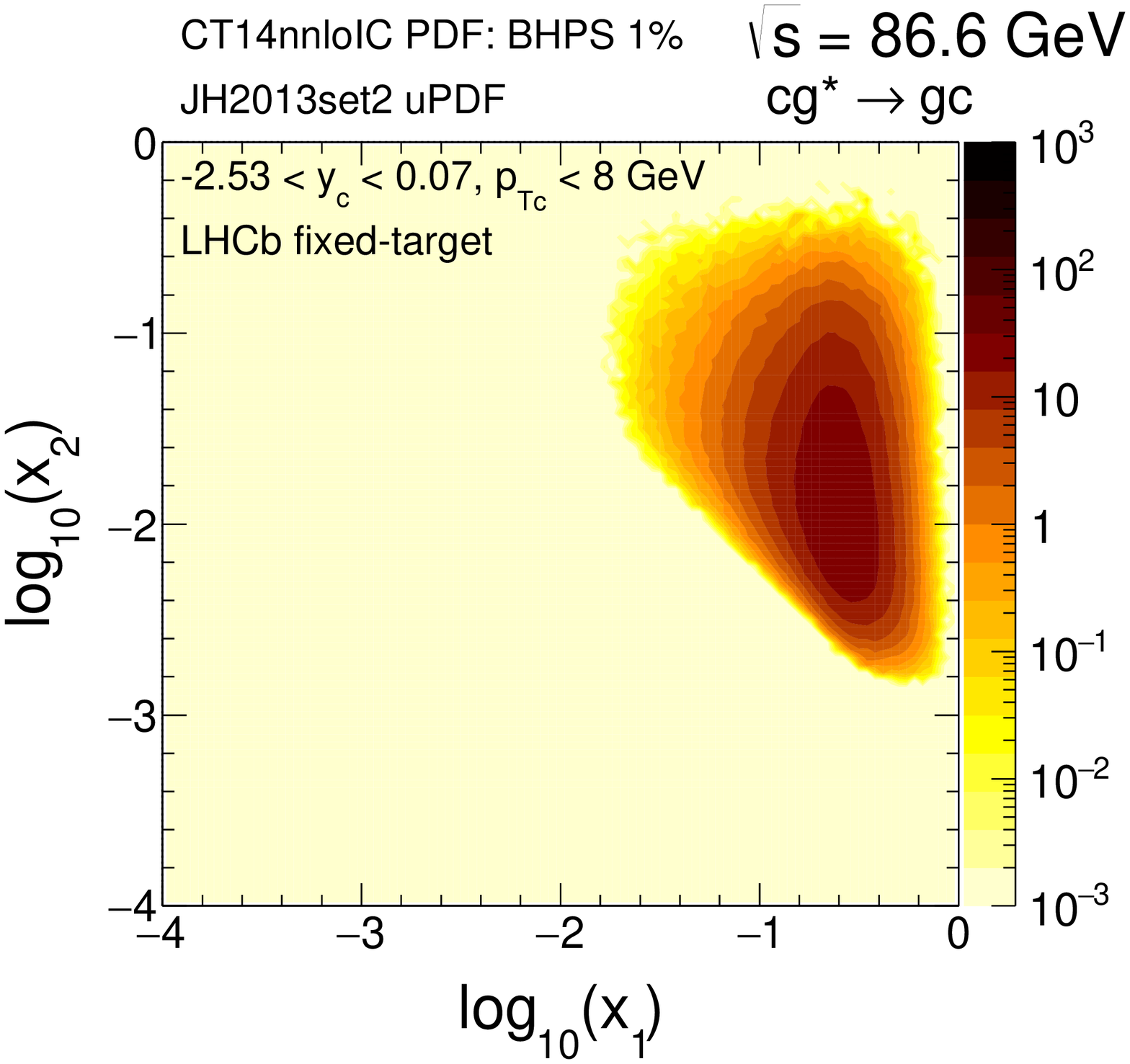}}
\end{minipage}
\begin{minipage}{0.33\textwidth}
  \centerline{\includegraphics[width=1.0\textwidth]{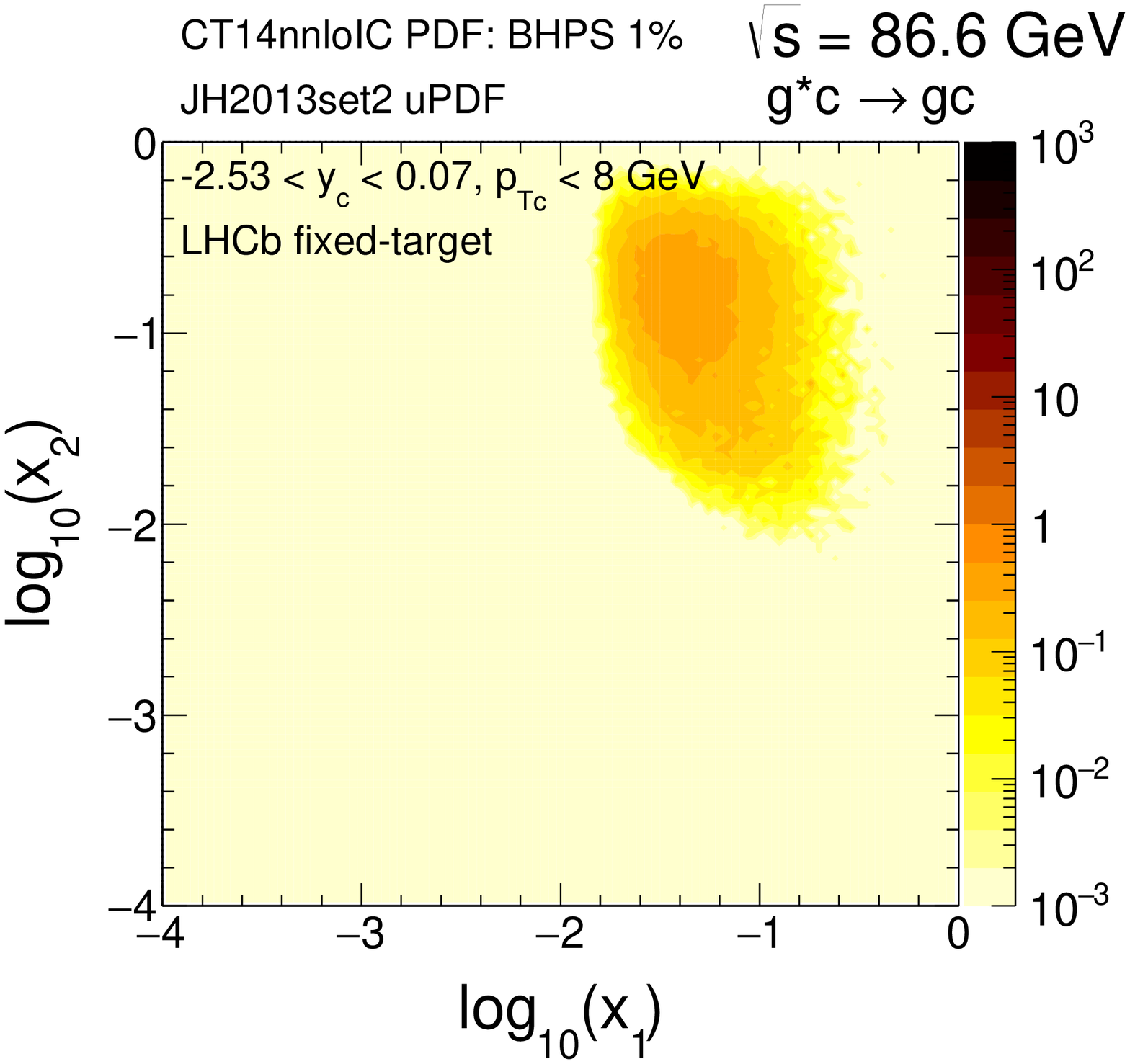}}
\end{minipage}
  \caption{
\small The two-dimensional distributions as a function of longitudinal momentum fractions $\log_{10}(x_{1})$ and $\log_{10}(x_{2})$ probed in the LHCb fixed-target acceptance. The cross sections for $cg^* \to gc$ (left panel) and $g^*c \to gc$ (right panel) mechanisms are shown separately. Details are specified in the figure.
}
\label{fig:3}
\end{figure}

In the present paper, we wish to concentrate mostly on the mechanism associated with the intrinsic charm concept in proton which was not discussed so far
in the context of the LHCb fixed-target charm data. Before we go to the main results we wish to visualize qualitatively the kinematics behind the considered production mechanism. In Fig.~\ref{fig:3} we present double differential parton-level cross section for charm quark as a function of longitudinal momentum fractions $\log_{10}(x_{1})$ and $\log_{10}(x_{2})$ carried by the incident partons for $cg^* \to gc$ (left panel) and $g^*c \to gc$ (right panel) mechanisms. Here we impose on the produced charm quark/antiquark cuts relevant for the LHCb fixed-target mode. We clearly see that one could probe here the unintegrated gluon distributions at rather intermediate $x$-values with maximum of the cross section around $10^{-2}$ and simultaneously the collinear charm distributions at very large $x$-values, larger than $10^{-1}$. This so far unexplored kinematical domain is of course very interesting and could help to constrain both, the unintegrated gluon and the collinear charm distributions in these exotic limits.

\begin{figure}[!h]
\begin{minipage}{0.47\textwidth}
  \centerline{\includegraphics[width=1.0\textwidth]{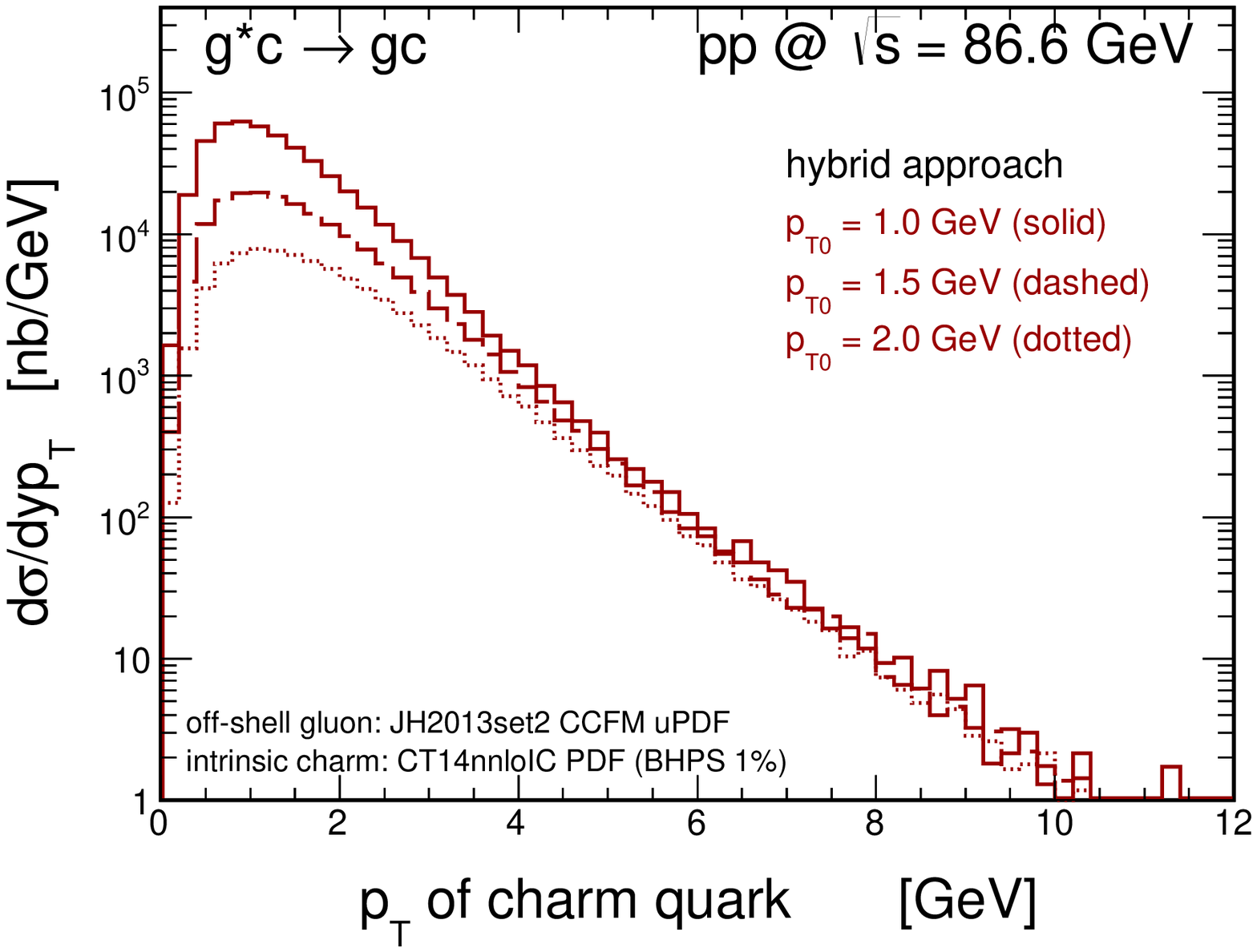}}
\end{minipage}
\begin{minipage}{0.47\textwidth}
  \centerline{\includegraphics[width=1.0\textwidth]{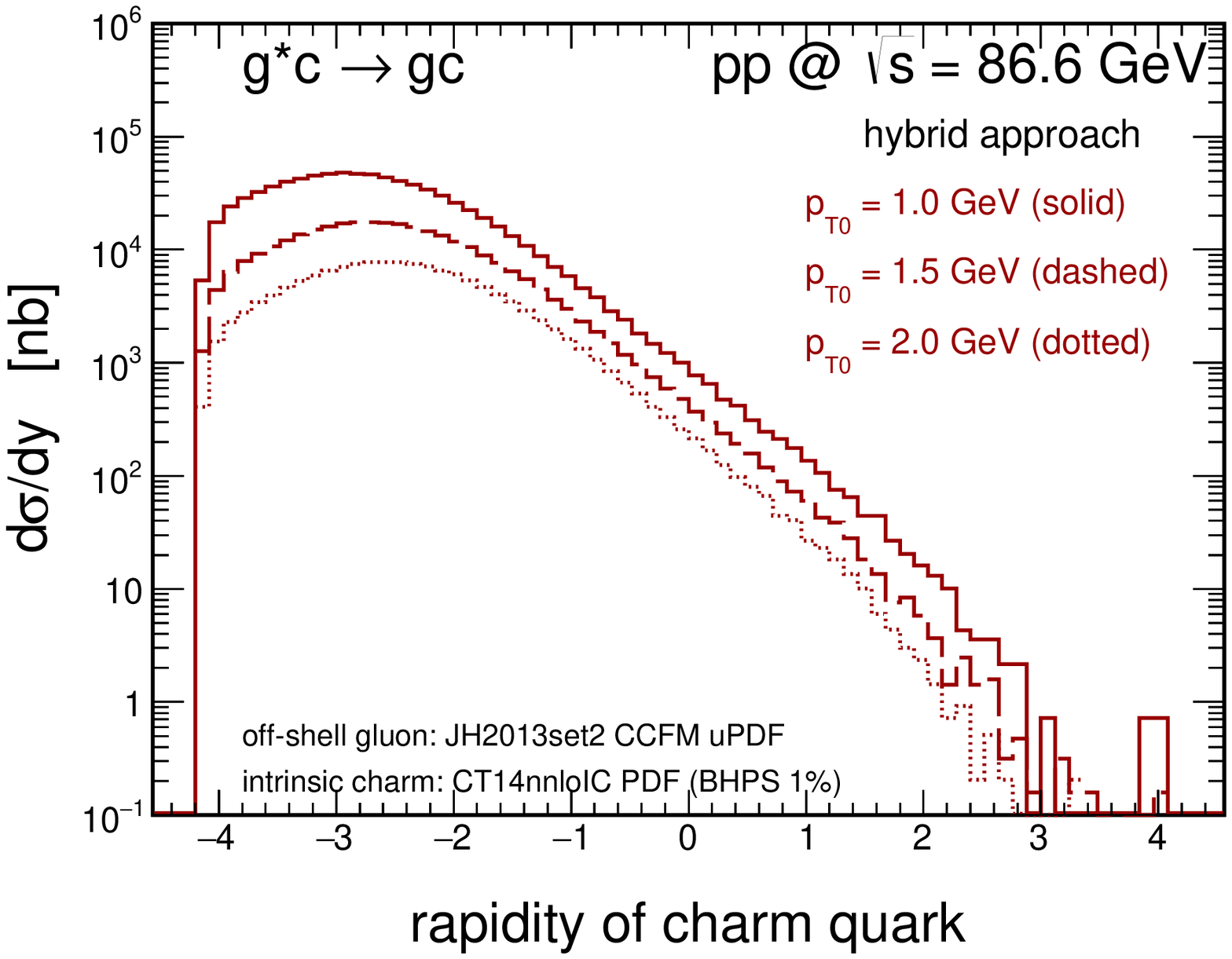}}
\end{minipage}
  \caption{
\small The transverse momentum (left) and rapidity (right) distributions of charm quark
for $pp$-collisions at $\sqrt{s} = 86.6$ GeV for the $g^*c\to gc$ mechanism calculated in the hybrid model assuming  
the BHPS 1\% model for intrinsic charm and taking the JH2013set2 CCFM gluon uPDF. Here results for three different values of $p_{T0}$ are shown. Details are specified in the figure.
}
\label{fig:5}
\end{figure}

As was already introduced in the previous section (and also discussed in Ref.~\cite{Maciula:2020dxv}), our model for calculation of the intrinsic charm contribution to the charm production cross section depends on a free parameter $p_{T0}$ used for a regularization of the cross section. From Fig.~\ref{fig:5} we see that the predictions for charm quark transverse momentum (left panel) and rapidity (right panel) distributions are very sensitive to the choice of this parameter, especially, at small charm quark transverse momenta, which also affects the rapidity spectrum. At larger transverse momenta, the sensitivity disappears. In the numerical studies below the three different values $p_{T0} = 1.0$, $1.5$, and $2.0$ GeV were used and examined against the LHCb data sets.

\begin{figure}[!h]
\begin{minipage}{0.47\textwidth}
  \centerline{\includegraphics[width=1.0\textwidth]{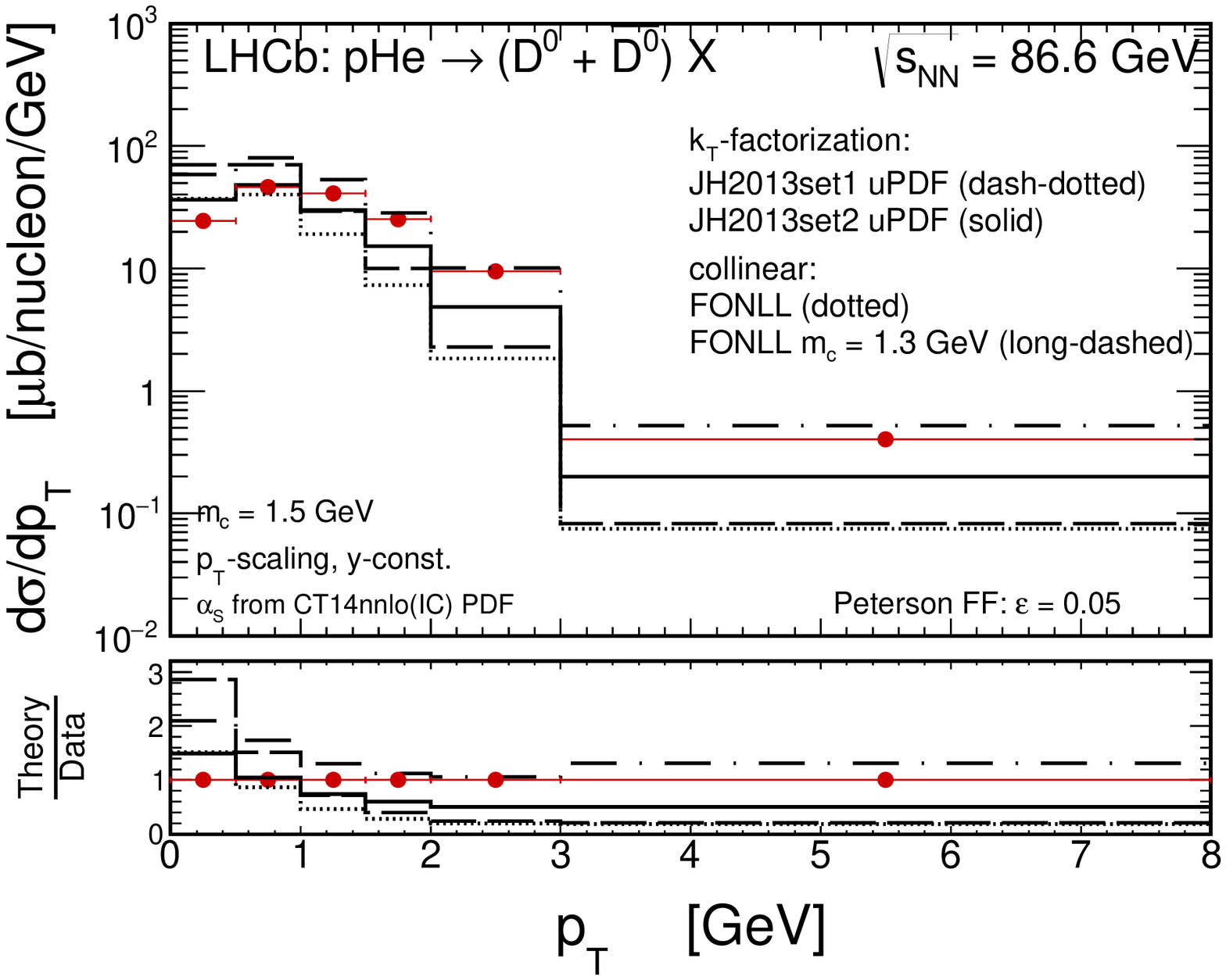}}
\end{minipage}
\begin{minipage}{0.47\textwidth}
  \centerline{\includegraphics[width=1.0\textwidth]{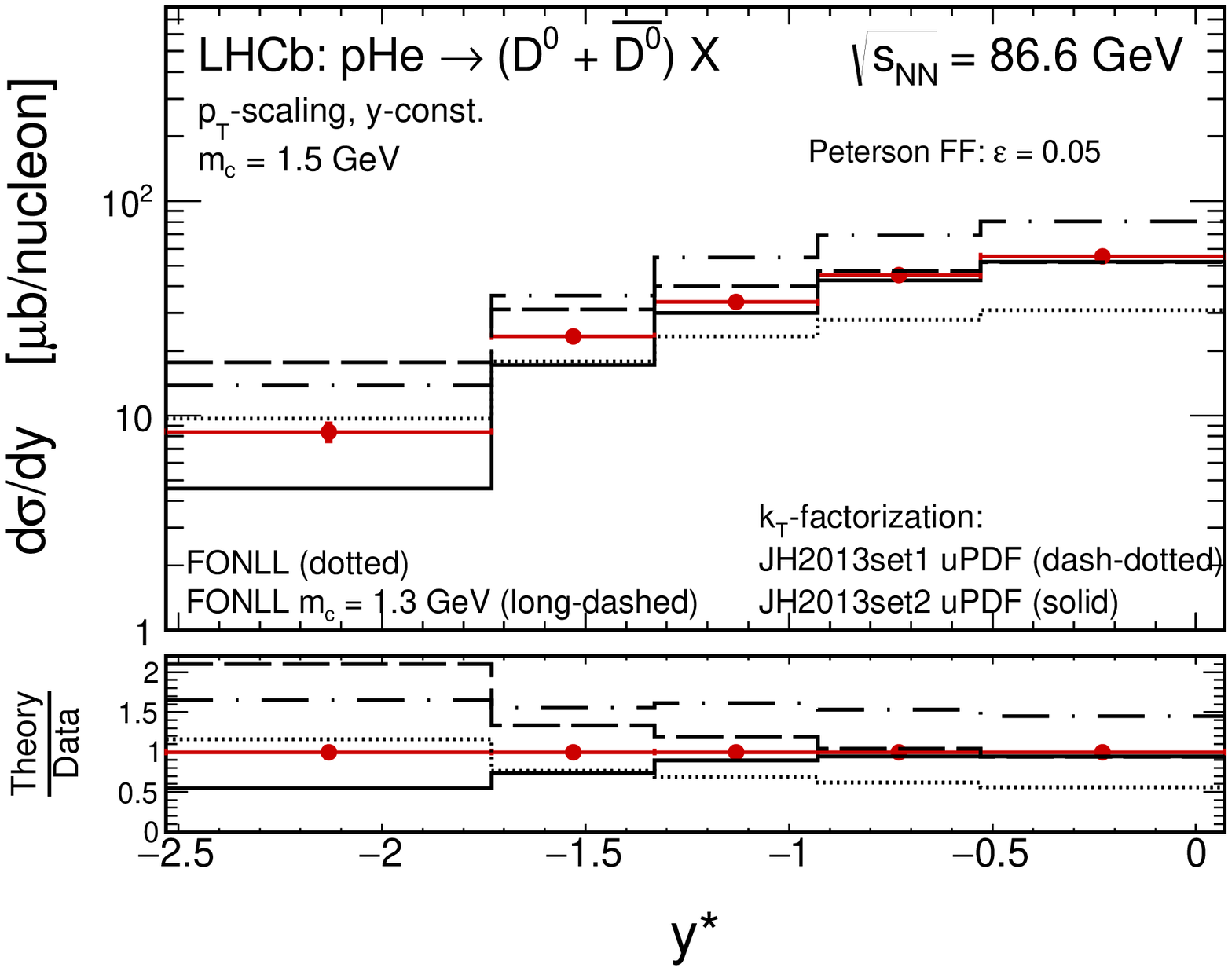}}
\end{minipage}
  \caption{
\small The transverse momentum (left) and rapidity (right) distributions of $D^{0}$ meson (plus $\overline{D^{0}}$ antimeson)
for $p+^4\!\mathrm{He}$ collisions together with the LHCb data \cite{Aaij:2018ogq}. Here results of the standard calculations of the $c\bar c$-pair production are shown without any intrinsic charm component. Details are specified in the figure.
}
\label{fig:6}
\end{figure}

\begin{figure}[!h]
\begin{minipage}{0.47\textwidth}
  \centerline{\includegraphics[width=1.0\textwidth]{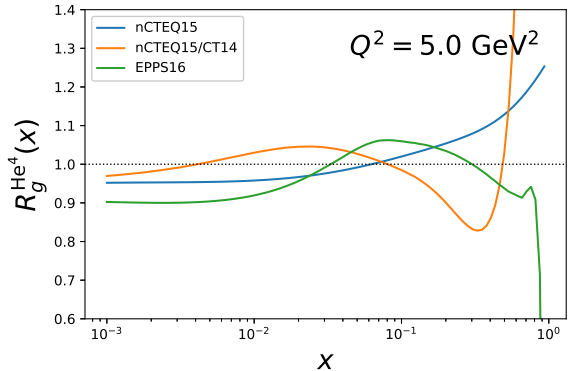}}
\end{minipage}
  \caption{
\small The ratio of nuclear to nucleon gluon distributions for three
different sets from the literature.
}
\label{fig:nuclear-effects}
\end{figure}

Now we start presentation of our numerical results for $D^{0} + \overline{D^{0}}$ meson production in fixed-target $p+^4\!\mathrm{He}$ collisions at $\sqrt{s_{NN}}=86.6$ GeV, measured for the first time very recently by the LHCb collaboration \cite{Aaij:2018ogq}. The experimental cross sections are divided by the number of nucleons and are compared below with the theoretical results for $pp$-scattering. Nuclear effects in the case of the
$p+^4\!\mathrm{He}$ interactions and for considered kinematical range
are expected to be negligible, which was checked and explicitly shown in
Ref.~\cite{Aaij:2018ogq} (see Fig.~4 therein).
Fig.~\ref{fig:nuclear-effects} shows the ratio of nuclear ($^{4}He$)
to nucleon gluon distributions for a typical for our reaction, rather low, factorization 
scale $\mu^2$ = 5 GeV$^2$. The effect is rather small and different models
give rather contradictory results. Therefore in the following we shall
neglect the nuclear effects completely. Below we show in addition the theory/data ratio.
We start with the results for the standard $c\bar c$-pair production mechanism calculated both, in the framework of the $k_{T}$-factorization approach with off-shell initial state partons and with the CCFM uPDFs, and in the NLO collinear approach within the FONLL framework.

In Fig.~\ref{fig:6} we present differential distributions of $D^{0}$ meson as a function of c.m.s. rapidity (left panel) and transverse momentum (right panel) for $p+^4\!\mathrm{He}$ collisions together with the LHCb experimental data points \cite{Aaij:2018ogq}. A detailed phenomenological study of the data set within the $k_{T}$-factorization approach has been already performed by one of us in Ref.~\cite{Maciula:2020cfy}. There different models of unintegrated gluon densities were used and tested against the LHCb data. It was found that, in principle, the calculations based on the $k_{T}$-factorization with the CCFM unintegrated gluon densities are able to describe the data in a satisfactory manner. Here we repeat a part of the calculations presented there and use two different sets of the gluon CCFM uPDFs: the JH-2013-set1 and JH-2013-set2 fits \cite{Hautmann:2013tba}, however, the results obtained here may slightly differ from those reported in Ref.~\cite{Maciula:2020cfy} since the choice of the strong coupling is not the same. Here and in the following, we keep the $\alpha_{S}(\mu^2)$ at NNLO in order to be consistent with the $\alpha_{S}(\mu^2)$ grids present in the CT14nnloIC PDFs.   
Predictions obtained within both gluon uPDF models significantly differ from each other. The JH-2013-set1 results seem to slightly overestimate the LHCb data points, especially in the region of small meson transverse momentum which also affects the rapidity distribution. On the other hand, the JH-2013-set2 model leads to a smaller cross section, providing a good description of the data at small meson transverse momentum and at mid-rapidities, but fails at larger meson $p_{T}$'s as well as in the backward rapidity region. Both, the JH-2013-set1 and JH-2013-set2 gluon uPDFs are determined from high-precision DIS measurements, including experimental and theoretical uncertainties. However, the JH-2013-set1 is determined from the fit to inclusive $F_2$ data only while the JH-2013-set2 is determined from the fit to both $F^{(\mathrm{charm})}_2$ and $F_2$ data. Drawing a definite conclusions is therefore not easy. For comparison we show also predictions of the FONLL framework for two different values of charm quark mass. Within the default set $m_{c}=1.5$ GeV, FONLL prediction significantly underestimates the LHCb data and leads to the cross sections smaller than in the case of the $k_{T}$-factorization approach.

\begin{figure}[!h]
\begin{minipage}{0.47\textwidth}
  \centerline{\includegraphics[width=1.0\textwidth]{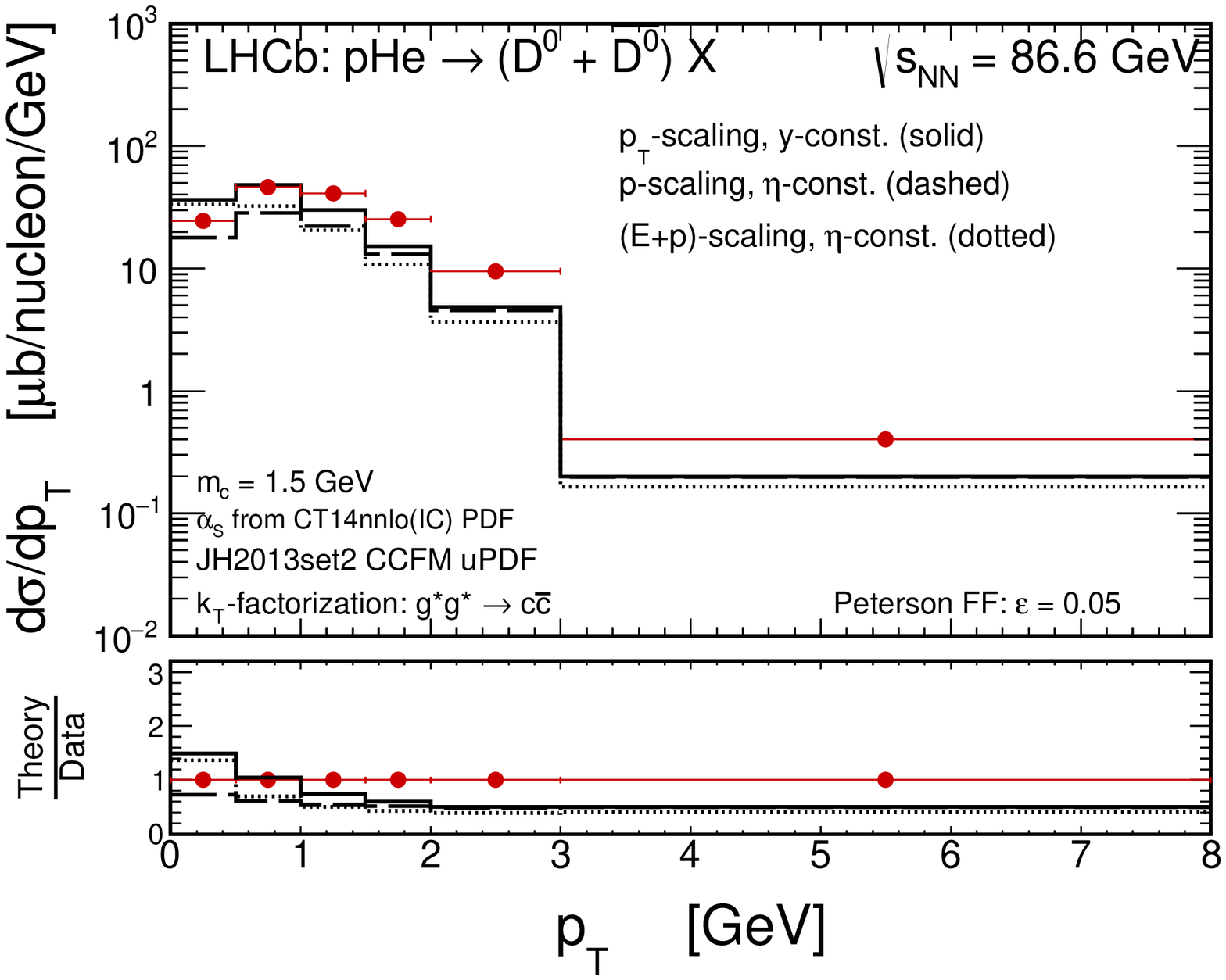}}
\end{minipage}
\begin{minipage}{0.47\textwidth}
  \centerline{\includegraphics[width=1.0\textwidth]{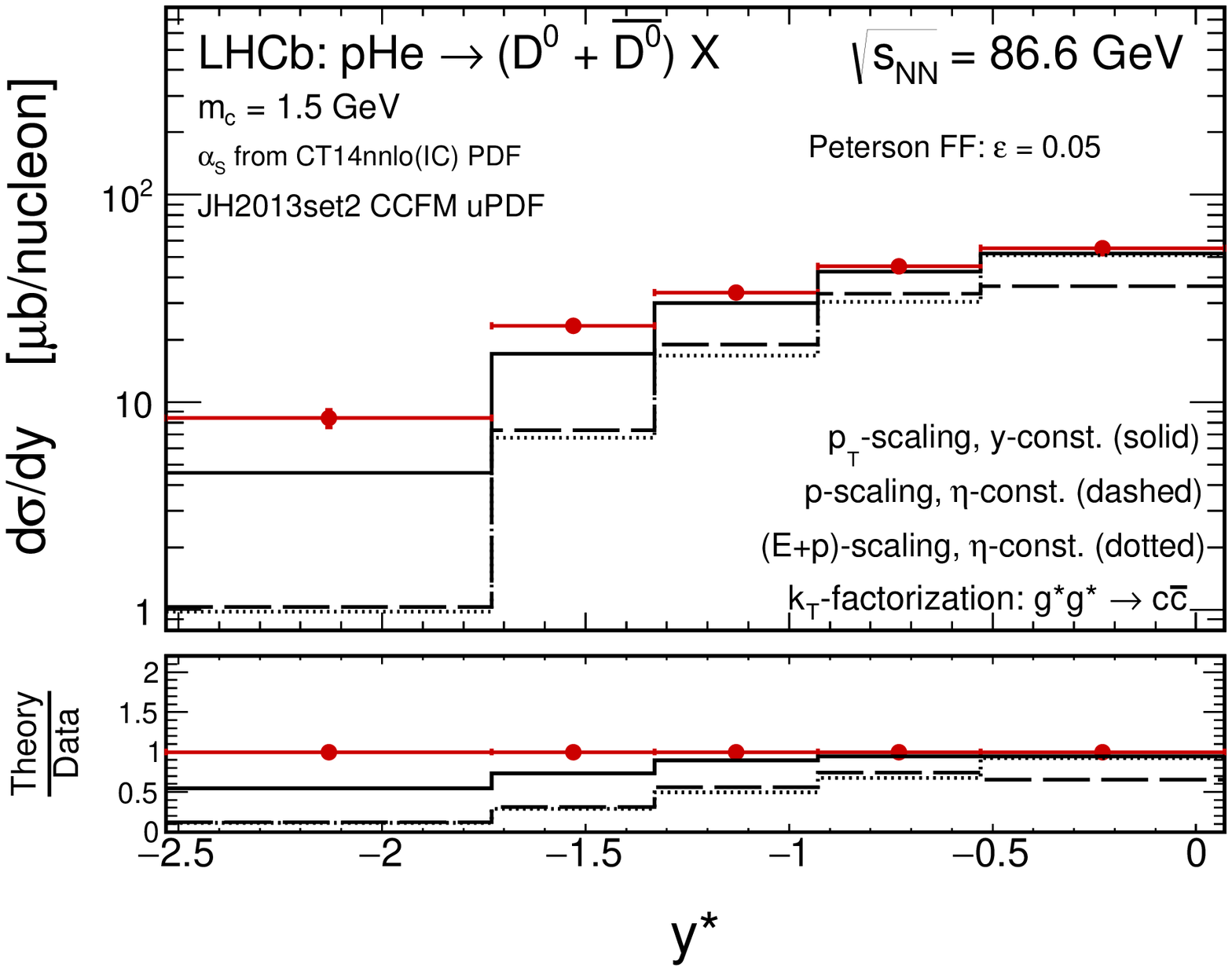}}
\end{minipage}
  \caption{
\small The transverse momentum (left) and rapidity (right) distributions of $D^{0}$ meson (plus $\overline{D^{0}}$ antimeson)
for $p+^4\!\mathrm{He}$ collisions together with the LHCb data \cite{Aaij:2018ogq}. Here results for three different approaches used in the procedure of the fragmentation of charm quark are shown. Details are specified in the figure.  
}
\label{fig:7}
\end{figure}

Now we wish to discuss the aspect of above calculations that was not analysed in Ref.~\cite{Maciula:2020cfy}. Namely, we wish to pay more attention to the details of the fragmentation procedure. In the following we show for the first time the result of the $\eta_D = \eta_c$ prescription with scaling variable defined as follows: $p_H = z p_q$ (momentum scaling) or $(E+p)_H = z (E+p)_q$ (light-cone scaling). In Fig.~\ref{fig:7} we compare results obtained within these two prescriptions to those calculated within the standard procedure with unchanged rapidity. Clearly, a significant difference between the standard and the low-energy model appears at small meson transverse momenta ($p_{T} < 2$ GeV) which in consequence leads to a quite different rapidity distribution. It seems that in the backward rapidity region details of the fragmentation procedure are of a special importance
and could lead to different conclusions about quality of the theoretical predictions and understanding of the experimental data.

As it is shown in Fig.~\ref{fig:8}, when applying the low-energy model for fragmentation with the light-cone momentum scaling the overall picture presented in Fig.~\ref{fig:6} is changed. Now, the $k_{T}$-factorization predictions with the JH-2013-set1 gluon uPDFs lie much closer to the LHCb data points giving their satisfactory description while the calculation with the JH-2013-set2 uPDF model underestimates measured cross sections (see top panels) . Different fragmentation procedures lead to different scenarios. The new scenario presented here might be more accurate than the standard one in the kinematics probed by the LHCb in its fixed-target mode. 
The fact that within this model the LHCb data favour more results for the JH-2013-set1 gluon uPDF than for the JH-2013-set2 gluon uPDF is a bit unexpacted, since the latter is determined from the fit to both $F^{(\mathrm{charm})}_2$ and $F_2$ DIS data.

\begin{figure}[!h]
\begin{minipage}{0.47\textwidth}
  \centerline{\includegraphics[width=1.0\textwidth]{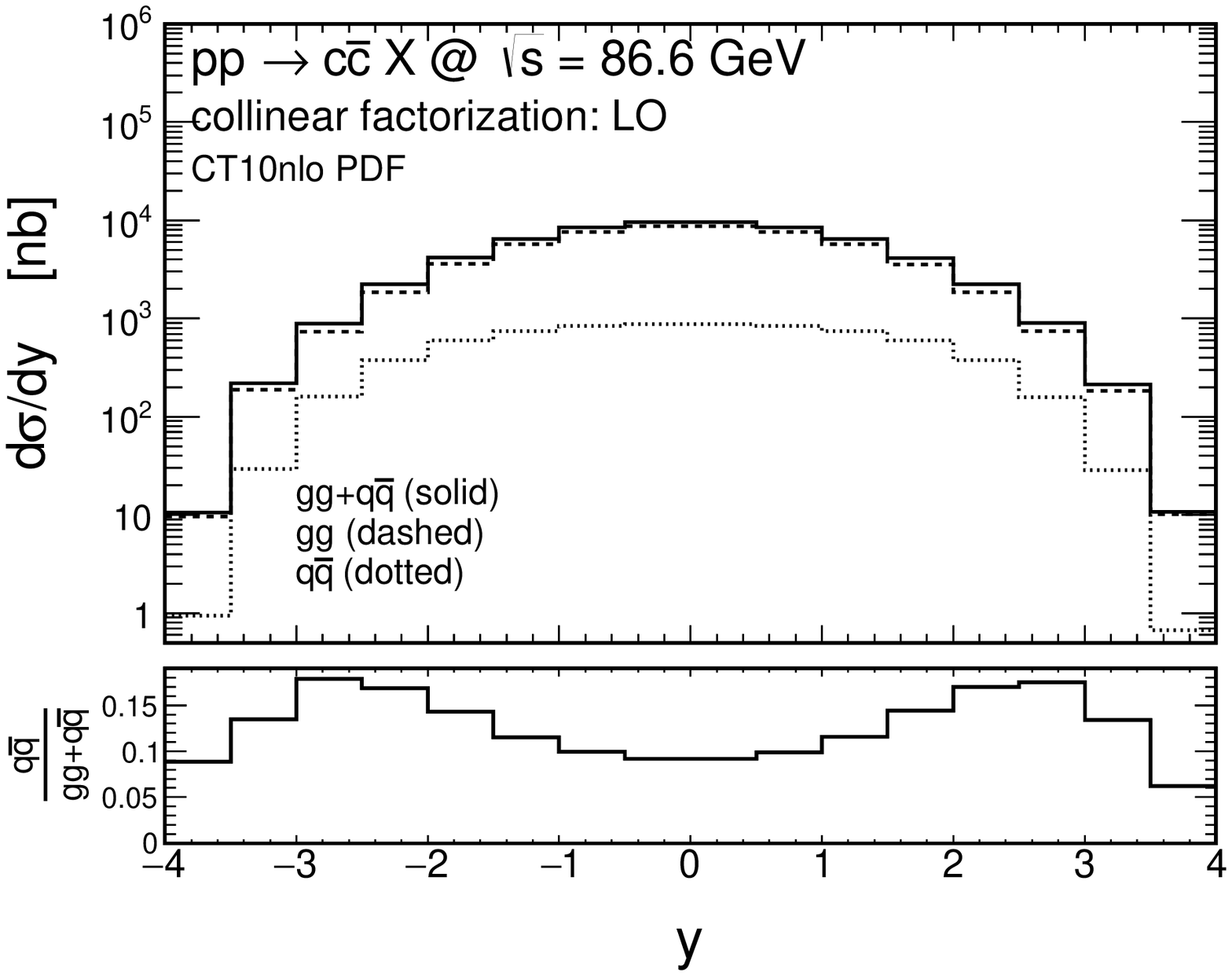}}
\end{minipage}
\begin{minipage}{0.47\textwidth}
  \centerline{\includegraphics[width=1.0\textwidth]{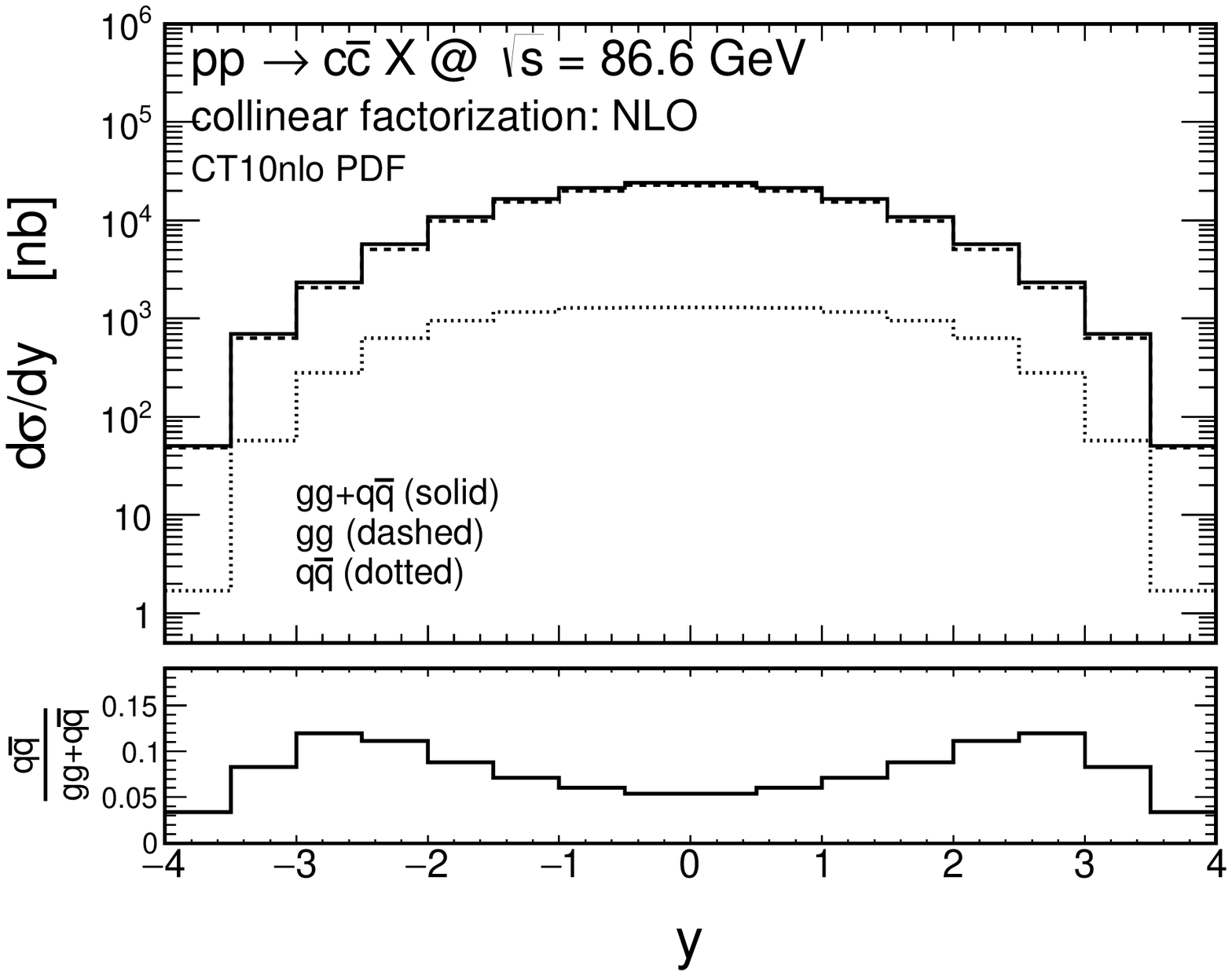}}
\end{minipage}
  \caption{
\small A direct comparison of the $gg$-fusion and the $q\bar{q}$-annihilation channels of the $c\bar{c}$-pair production at $\sqrt{s} = 86.6$ GeV for the LO (left panel)
and the NLO (right panel) collinear frameworks.
}
\label{fig:GGvsQQ}
\end{figure}

\begin{figure}[!h]
\begin{minipage}{0.47\textwidth}
  \centerline{\includegraphics[width=1.0\textwidth]{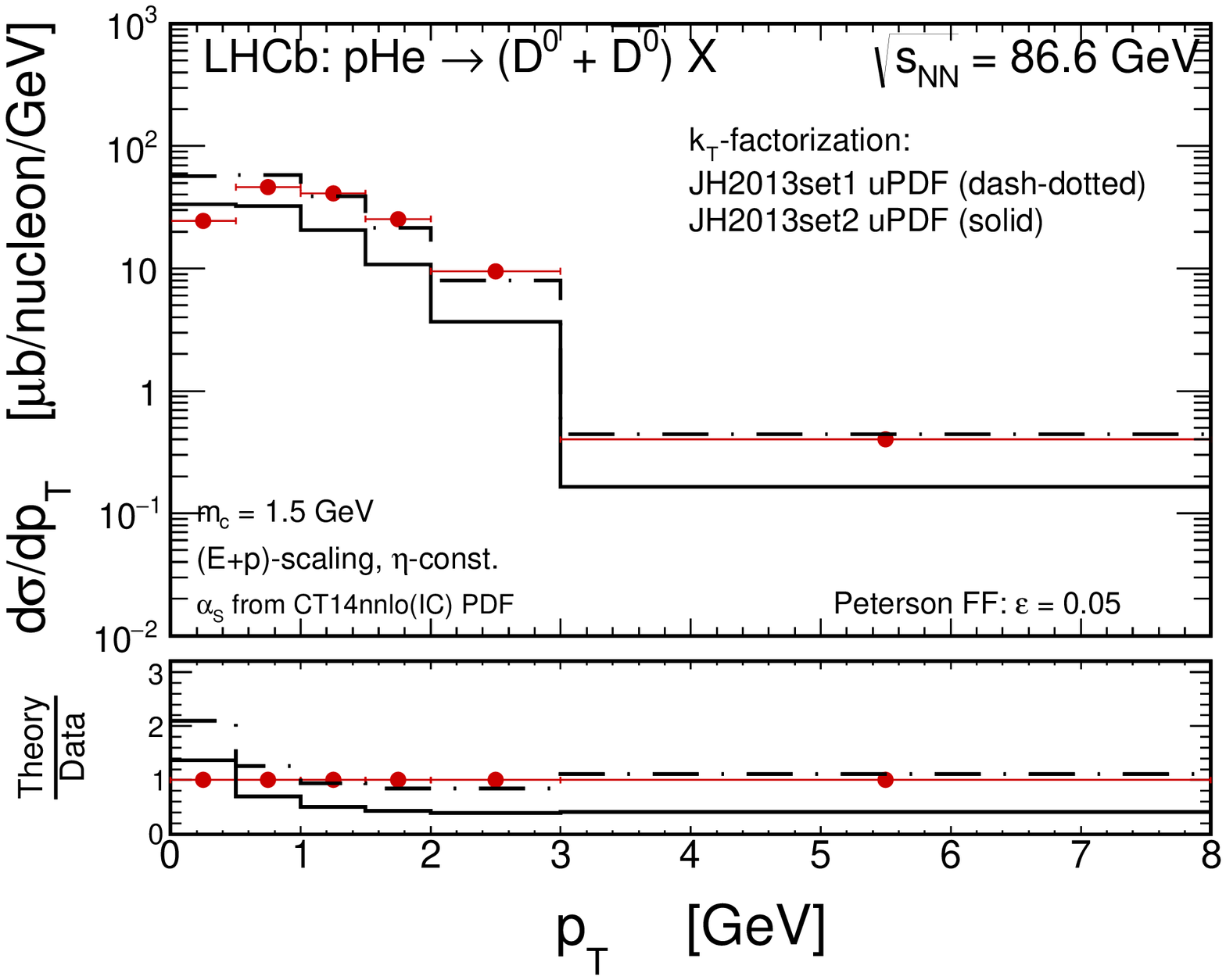}}
\end{minipage}
\begin{minipage}{0.47\textwidth}
  \centerline{\includegraphics[width=1.0\textwidth]{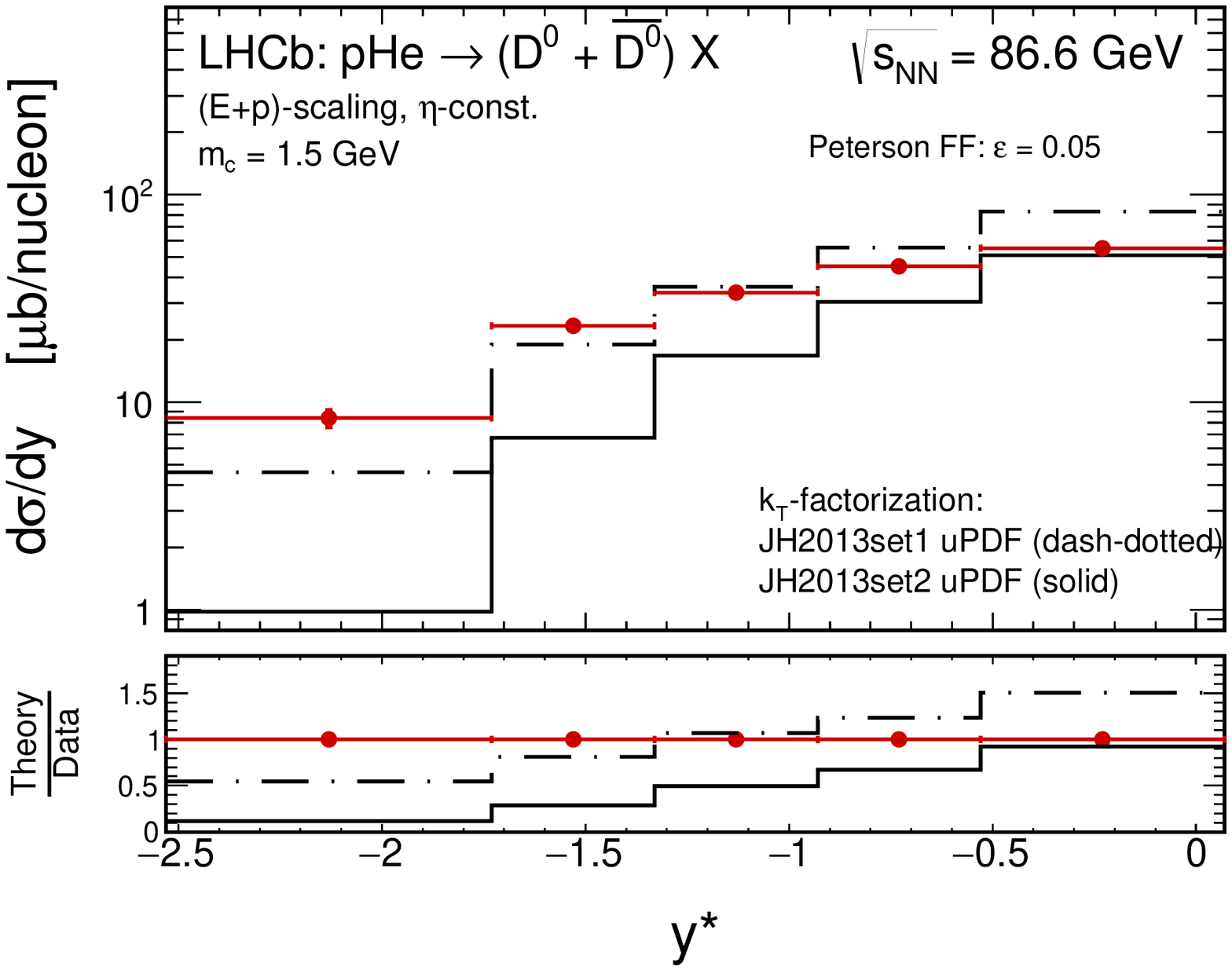}}
\end{minipage}
\begin{minipage}{0.47\textwidth}
  \centerline{\includegraphics[width=1.0\textwidth]{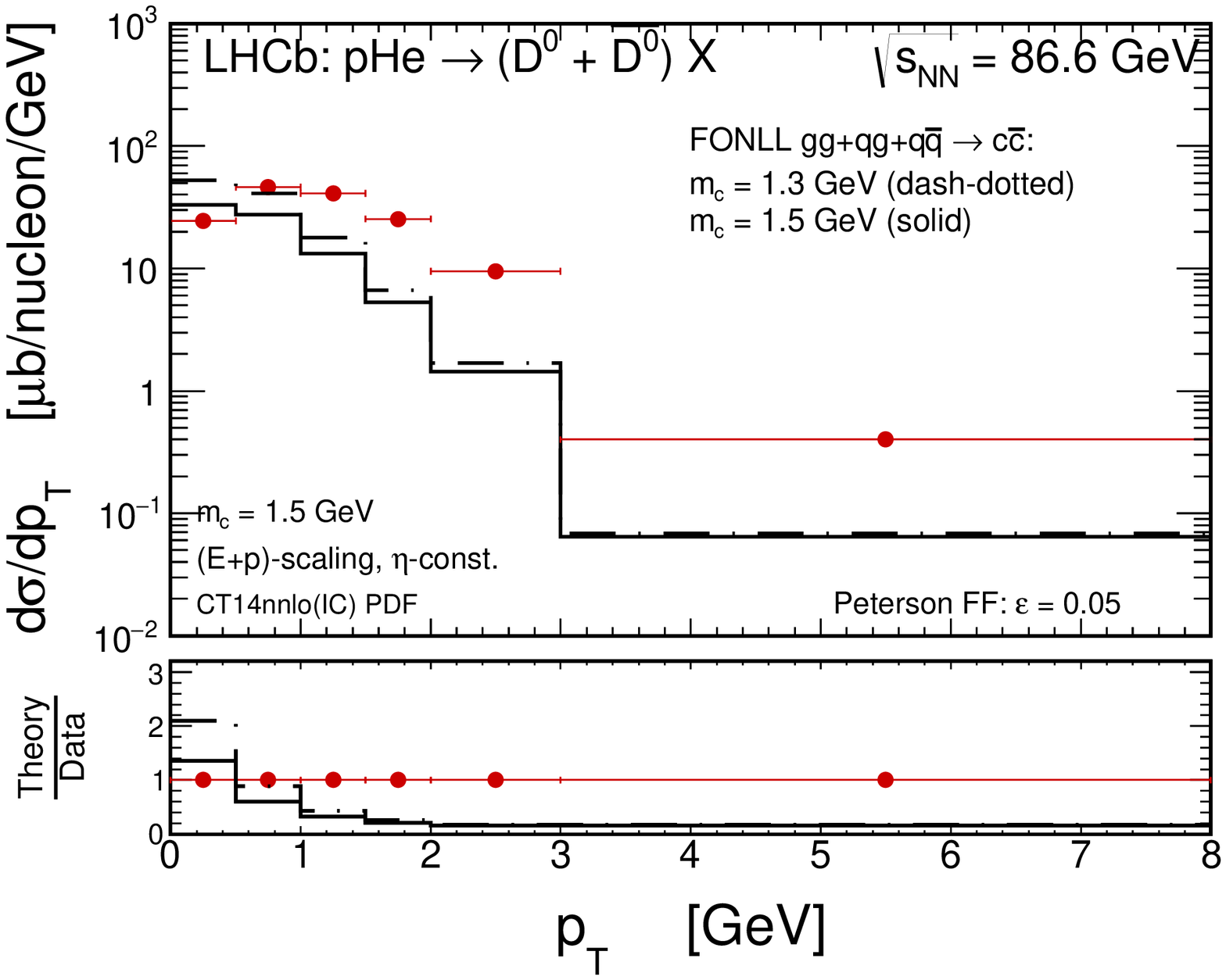}}
\end{minipage}
\begin{minipage}{0.47\textwidth}
  \centerline{\includegraphics[width=1.0\textwidth]{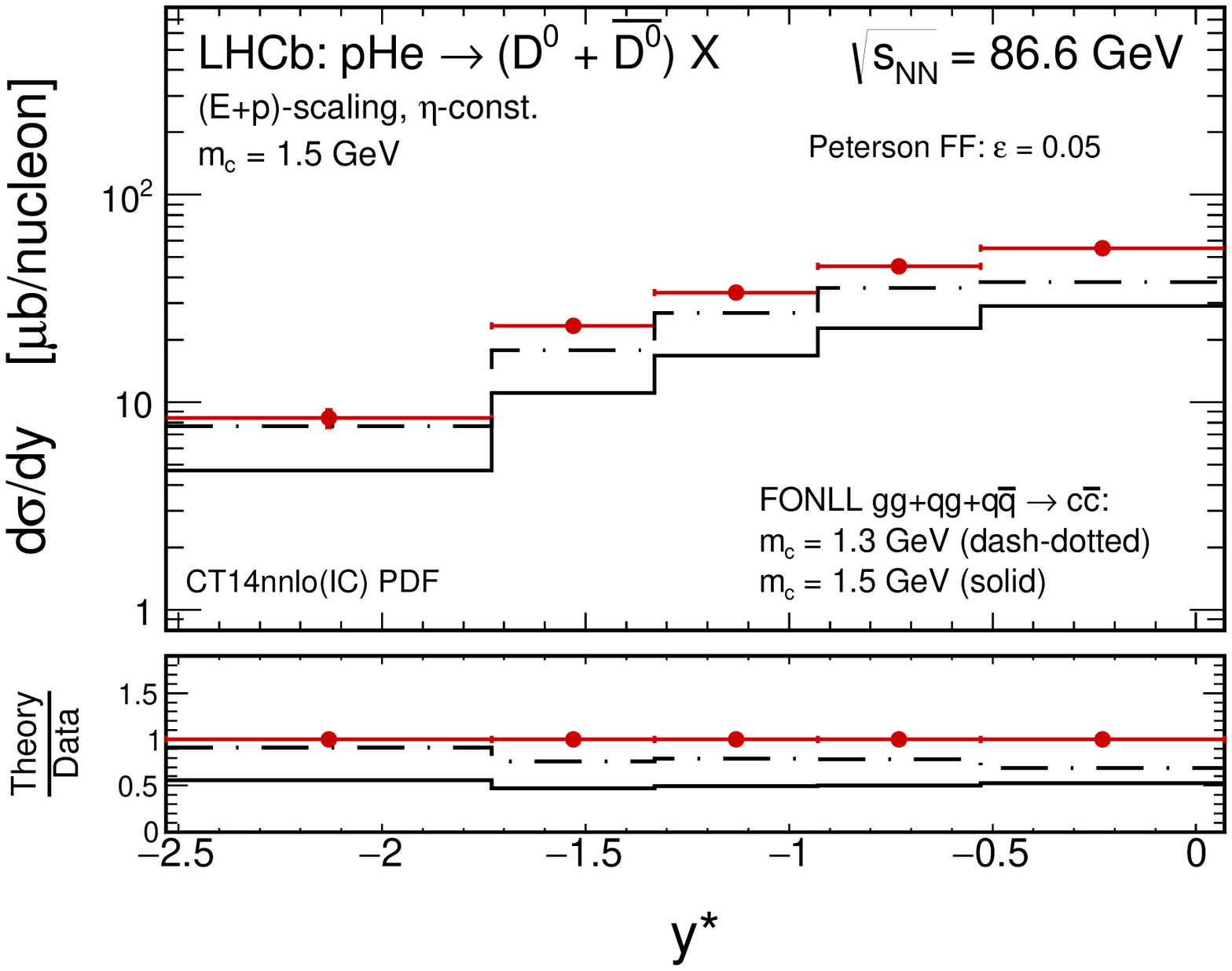}}
\end{minipage}
  \caption{
\small The transverse momentum (left) and rapidity (right) distributions of $D^{0}$ meson (plus $\overline{D^{0}}$ antimeson)
for $p+^4\!\mathrm{He}$ collisions together with the LHCb data
\cite{Aaij:2018ogq}. Here results of the standard calculations of
  the $c\bar c$-pair production obtained within the
  $k_{T}$-factorization framework for the JH2013set1 and JH2013set2
  gluon CCFM uPDFs (top panels) as well as within the collinear FONLL
  framework (bottom panels) are shown without any intrinsic charm
  component. Here light-cone momentum scaling with constant
quark pseudorapidity is used in the fragmentation procedure. 
Details are specified in the figure.
}
\label{fig:8}
\end{figure}

Considering the FONLL results plotted in Fig.~\ref{fig:8} (bottom panels) we see that the central predictions visibly underestimate
both data sets when $m_{c} = 1.5$ GeV is taken. The situation is slightly better for a smaller charm quark mass, i.e. $m_{c} = 1.3$ GeV, however, this
choice should be rather considered as an upper limit and not a central prediction. Keeping $m_{c} = 1.5$ GeV, we observe also that there is a sizeable difference between the FONLL prediction and the $k_{T}$-factorization result in the very backward rapidity region. We find the observation to be a direct consequence of the difference in the gluon densities used in both frameworks rather than the effect related to the subleading $q\bar{q} \to c\bar c$ production channels, part of which might be missing in the $k_{T}$-factorization approach\footnote{In principle, according to the $q\bar{q} \to c\bar c$ production mechanism, contributions coming from the light valence quark interactions might be limited in the $k_{T}$-factorization framework (in contrast to the sea-quark interactions).}. As we checked numerically at the considered $\sqrt{s} = 86.6$ GeV collision energy, for both the LO and the NLO collinear calculations the $gg$-fusion mechanism dominates over the $q\bar{q}$-annihilation channels even in the far forward/backward rapidity direction (see Fig.~\ref{fig:GGvsQQ}).

Starting from this point let us now follow a scenario in which the prediction obtained within the $k_{T}$-factorization framework with the JH-2013-set2 gluon uPDF, including fragmentation with the light-cone momentum scaling, is the point of reference suggested by the theory behind the model ingredients. Potentially, this picture leaves a room for missing charm production mechanisms at relatively large meson transverse momentum and in the region of backward meson rapidity. Here we propose the mechanism driven by intrinsic charm content of the nucleon as a possible source of a missing cross section reported within the present scenario.
It is interesting to check whether this idea may improve description of the LHCb data or not. It could also have a crucial meaning if the data could be used to constrain recent theoretical models of the intrinsic charm content of the nucleon.      

\begin{figure}[!h]
\begin{minipage}{0.47\textwidth}
  \centerline{\includegraphics[width=1.0\textwidth]{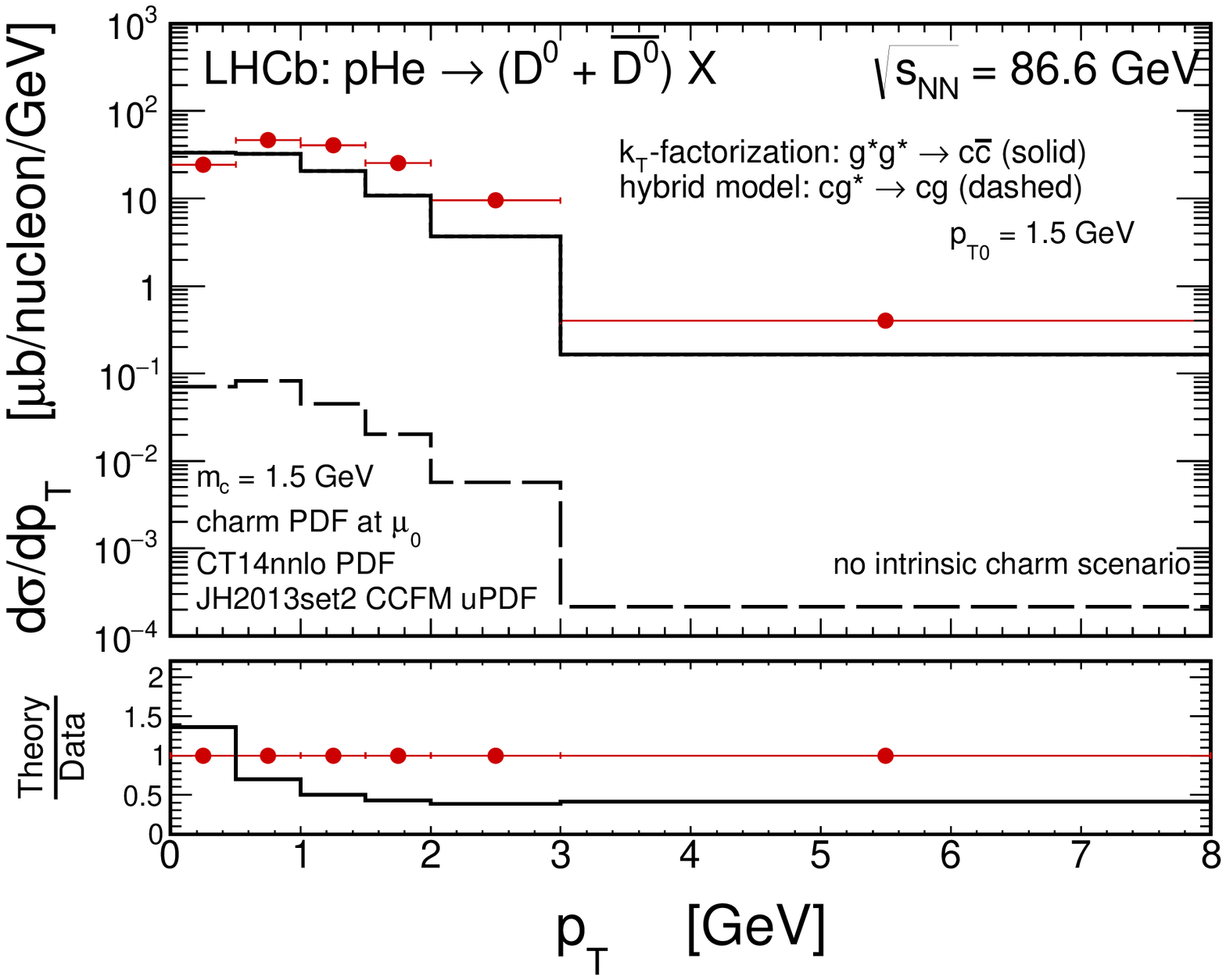}}
\end{minipage}
\begin{minipage}{0.47\textwidth}
  \centerline{\includegraphics[width=1.0\textwidth]{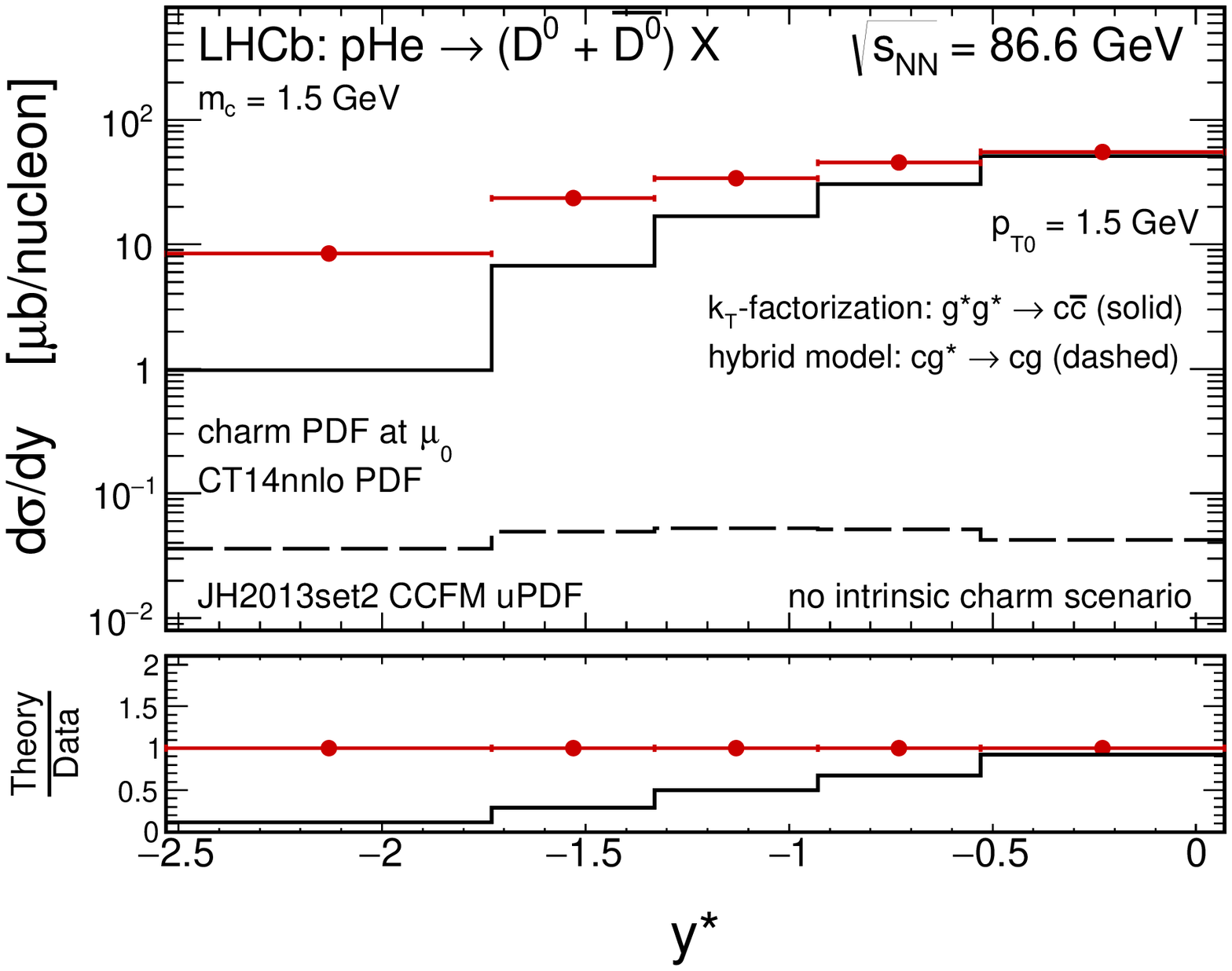}}
\end{minipage}
  \caption{
\small The transverse momentum (left) and rapidity (right) distributions of $D^{0}$ meson (plus $\overline{D^{0}}$ antimeson)
for $p+^4\!\mathrm{He}$ collisions together with the LHCb data
\cite{Aaij:2018ogq}. Here results of the standard calculations of the
$c\bar c$-pair production obtained with the $k_{T}$-factorization
framework and JH2013set2 gluon uPDF (solid histograms) as well as
results of the $g^*c \to gc$ mechanism obtained with the hybrid approach
and$g^*c \to gc$ with no intrinsic charm scenario (dashed histograms) are shown
separately. The $c$ or ${\bar c}$  PDF is taken here at the initial
scale $\mu_{0} = 1.3$ GeV. Other details are specified in the figure.  
}
\label{fig:9a}
\end{figure}

Before we start a presentation of the predicted intrinsic charm contribution to the differential distributions reported by the LHCb experiment we shall
check what is a contribution to the charm production cross section
coming from the $g^*c \to gc$ mechanism under the assumption that there
is no non-perturbative intrinsic $c {\bar c}$ content in the nucleon. So we test here the impact on the charm cross section of the charm quarks in the initial state produced only perturbatively,
during the QCD evolution. Here we also take the charm quark PDF at the
fixed factorization scale $\mu = 1.3$ GeV so we keep the PDF not much
evolved from the initial condition. The corresponding results of our
calculation are shown in Fig.~\ref{fig:9a}. Clearly, the $g^*c \to gc$
mechanism with no intrinsic charm leads to a smaller cross section than
the one predicted by the standard (gluon-gluon fusion or quark-antiquark
annihilation) $c\bar{c}$-pair production mechanism and cannot improve description of the LHCb open charm fixed-target data.

\begin{figure}[!h]
\begin{minipage}{0.47\textwidth}
  \centerline{\includegraphics[width=1.0\textwidth]{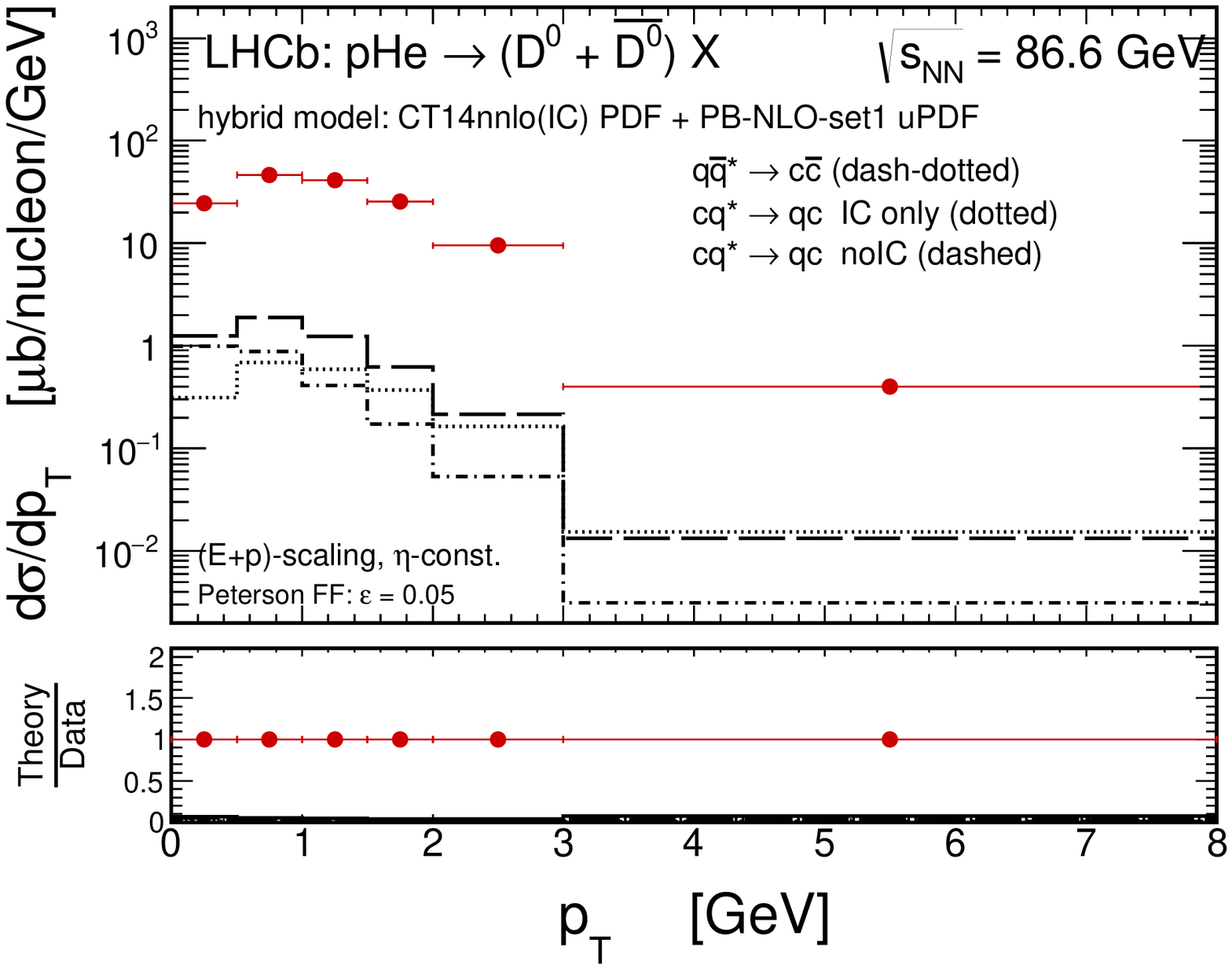}}
\end{minipage}
\begin{minipage}{0.47\textwidth}
  \centerline{\includegraphics[width=1.0\textwidth]{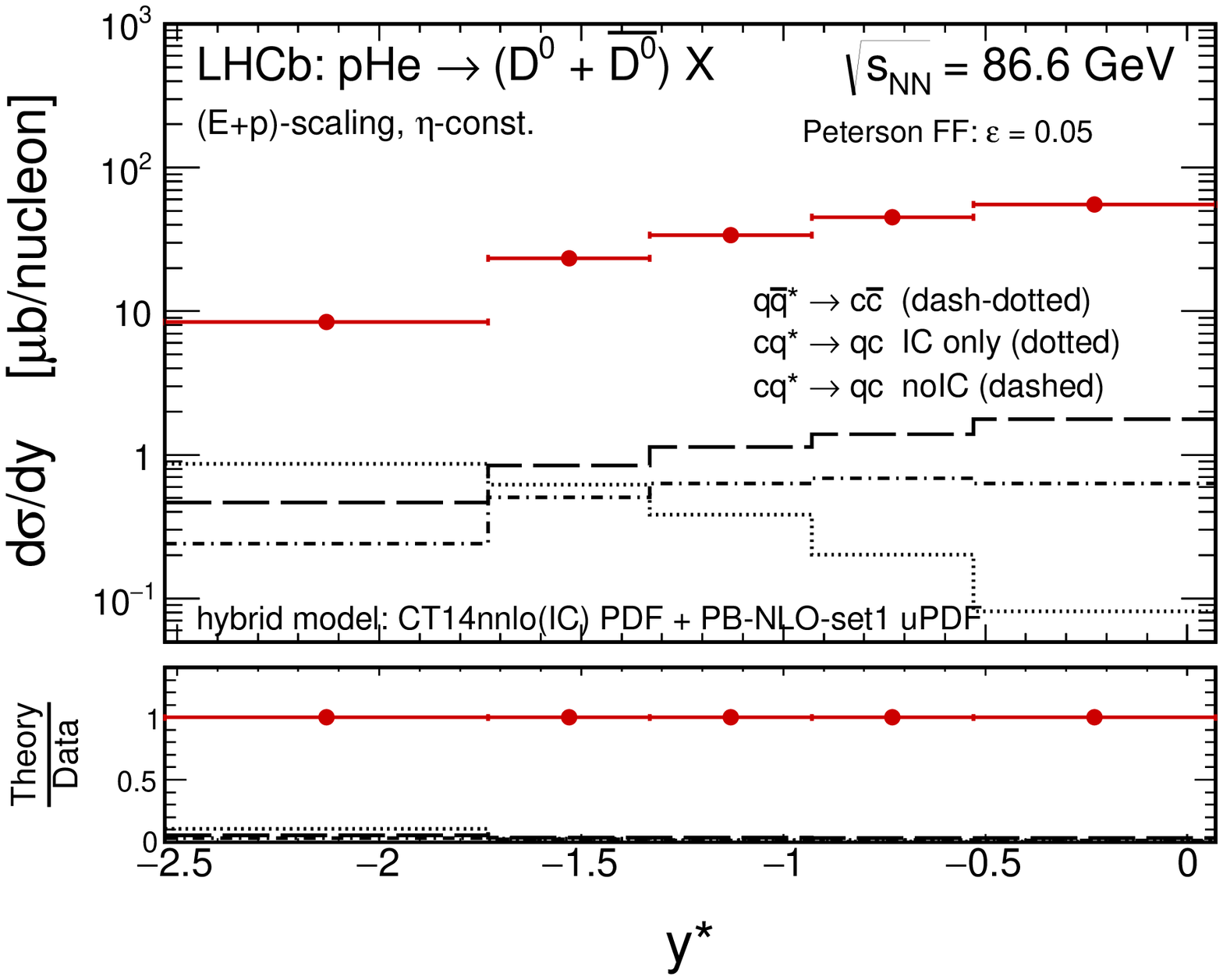}}
\end{minipage}
  \caption{
\small The transverse momentum (left) and rapidity (right) distributions of $D^{0}$ meson (plus $\overline{D^{0}}$ antimeson)
for $p+^4\!\mathrm{He}$ collisions together with the LHCb data
\cite{Aaij:2018ogq}. Here we compare results of two different subleading reactions driven by the interactions of quarks, namely the $q\bar{q}^* \to c\bar{c}$ (dash-dotted histograms) and the $cq^* \to qc$ (dashed and dotted histograms) mechanism. 
The results are obtained with the hybrid approach. Other details are specified in the figure.
}
\label{fig:9b}
\end{figure}

In the next step we examined also contributions coming from interactions of quarks only. As was shown e.g. in Ref.~\cite{Lipatov:2018oxm}, the mechanisms driven by $q\bar{q}$ and $cq$ interactions are competitive
with respect to the gluon-gluon fusion in the case of associated production of $Z$-boson and heavy flavoured jets at high energies. These contributions were found there to be important in the context of intrinsic charm studies. Here we present a similar analysis and include into numerical calculations the $q\bar{q}^{*} \to c \bar{c}$ (dash-dotted histograms) and the $cq^* \to qc$ (dashed and dotted histograms) mechanisms. The latter is calculated in two different ways: when only non-perturbative intrinsic charm quark PDF at the initial scale is taken into account (dotted histograms) as well as when only the perturbative charm quarks generated in the DGLAP evolution are used (dashed histograms). Here the PB-NLO-set1 \cite{BermudezMartinez:2018fsv} unintegrated quark distributions are used for the small-$x$ region. As can be concluded from Fig.~\ref{fig:9b}, at the energy and kinematics considered here, all three contributions are found to be subleading with respect to mechanisms with gluons in the initial state and can be safely neglected in the present study. Even in the very backward rapidity bin the quark initiated contributions are order of magnitude smaller than the leading mechanisms with gluons in the initial state.
  
\begin{figure}[!h]
\begin{minipage}{0.47\textwidth}
  \centerline{\includegraphics[width=1.0\textwidth]{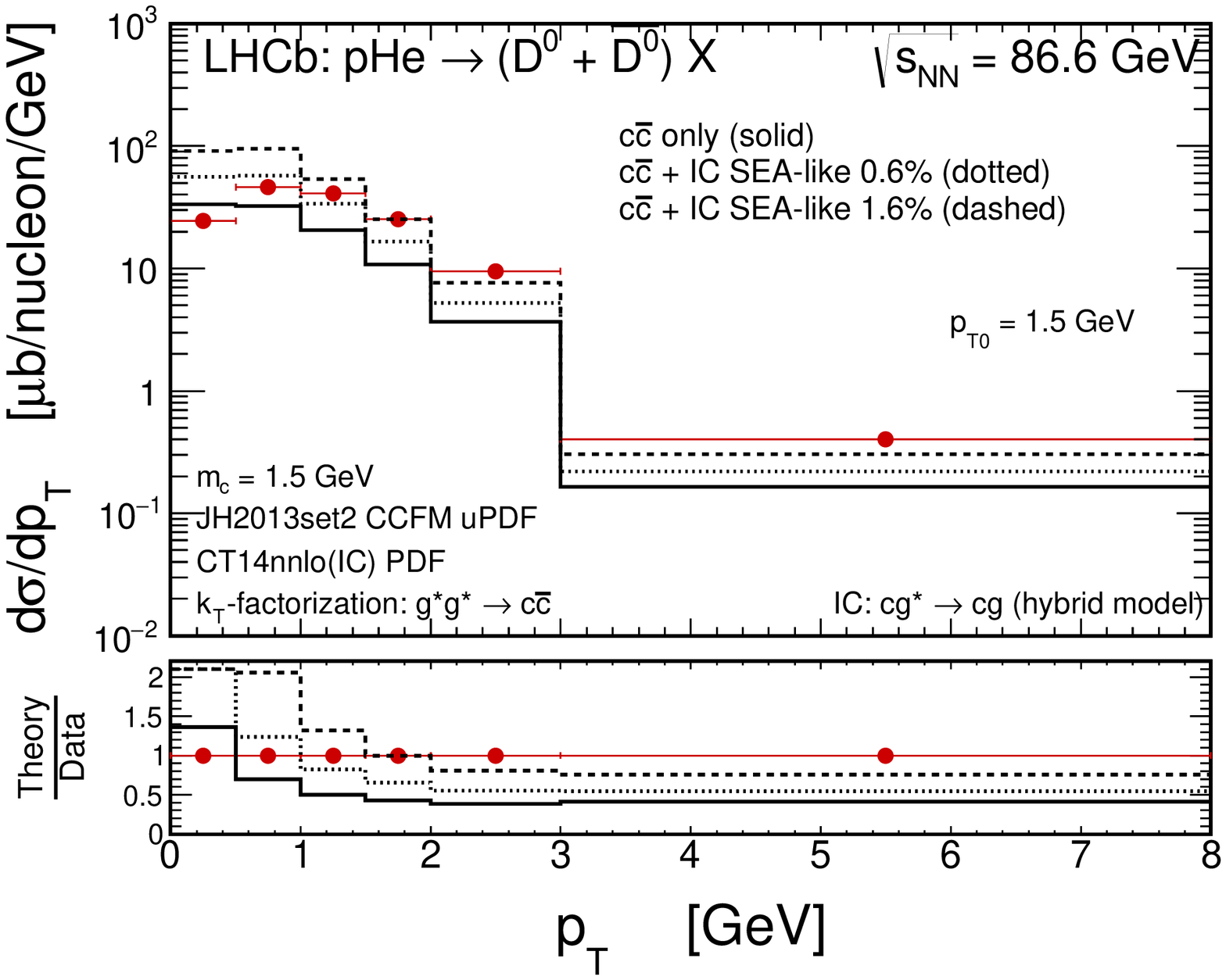}}
\end{minipage}
\begin{minipage}{0.47\textwidth}
  \centerline{\includegraphics[width=1.0\textwidth]{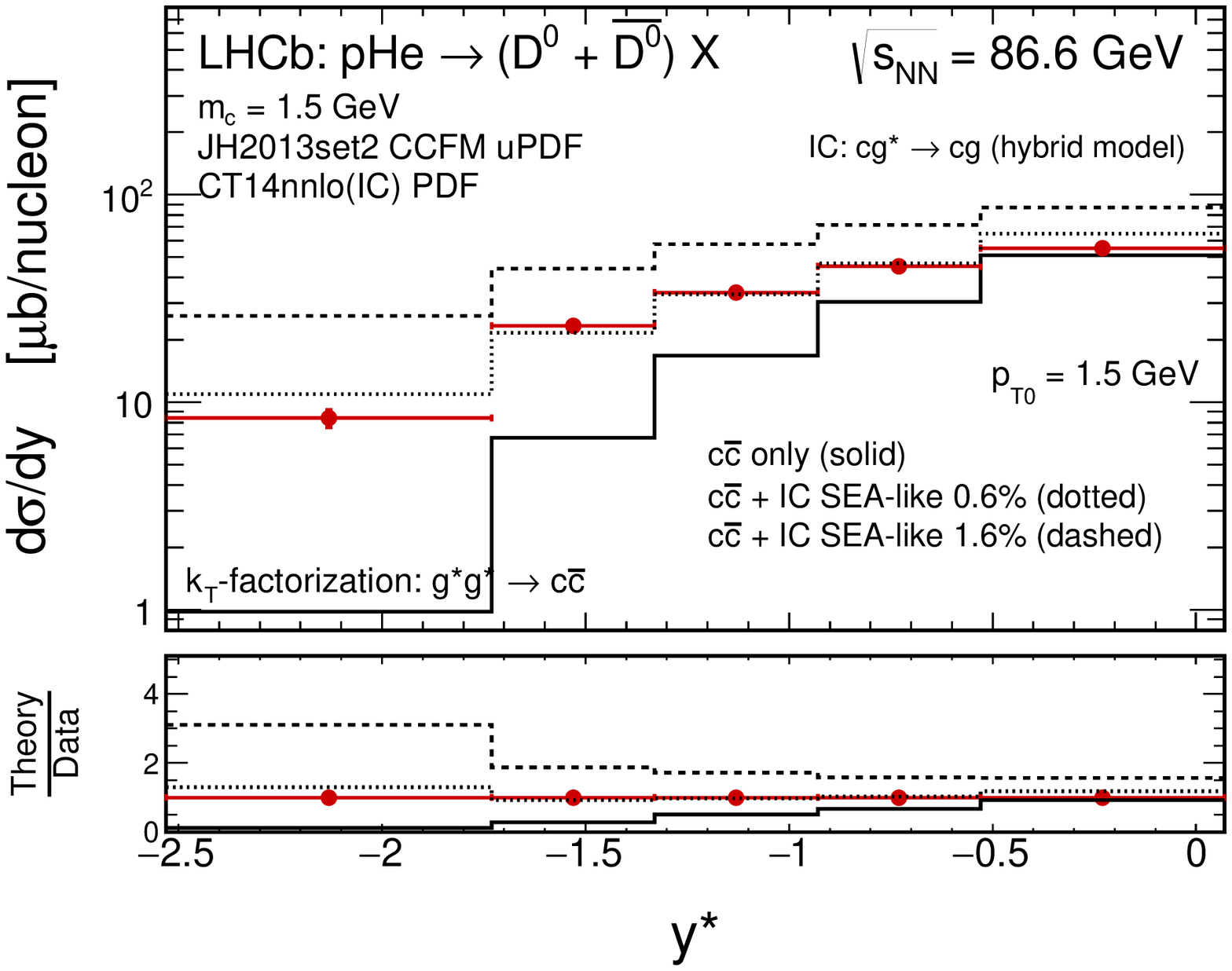}}
\end{minipage}
  \caption{
\small The transverse momentum (left) and rapidity (right) distributions of $D^{0}$ meson (plus $\overline{D^{0}}$ antimeson)
for $p+^4\!\mathrm{He}$ collisions together with the LHCb data \cite{Aaij:2018ogq}. Here results of the standard calculations of the $c\bar c$-pair production obtained with the $k_{T}$-factorization framework and JH2013set2 gluon uPDF are shown without any intrinsic charm component (solid histograms) as well as with different intrinsic charm contributions added (dotted, dashed and long-dashed histograms). Here the IC results correspond to the BHPS model. Other details are specified in the figure.  
}
\label{fig:9}
\end{figure}

Let us now show how the situation changes when the intrinsic charm content of the nucleon is taken into account. We start with predictions obtained within the 
SEA-like model for the intrinsic charm. In Fig.~\ref{fig:9} we again show the transverse momentum (left panel) and rapidity (right panel) distributions of $D^{0}$ meson. Here results of the standard calculations of the $c\bar c$-pair production are shown without any intrinsic charm component (solid histograms) as well as with different intrinsic charm contributions added. The dotted and dashed histograms correspond to the SEA-like model with the $P = 0.6\%$ and $P = 1.6\%$, respectively.
We see that the SEA-like models contribute more at small meson transverse momentum rather than at larger $p_{T}$-valus. The prediction of the SEA-like $P = 1.6\%$ overshoots
the LHCb data significantly at small meson transverse momentum. It also overestimates the rapidity spectrum in the whole considered range. The SEA-like $P = 0.6\%$
result lies closer to the data points, giving a reasonable description but still seems to overestimate the low-$p_{T}$ region and do not improve description of the data at the large-$p_{T}$ part of the distribution.   

\begin{figure}[!h]
\begin{minipage}{0.47\textwidth}
  \centerline{\includegraphics[width=1.0\textwidth]{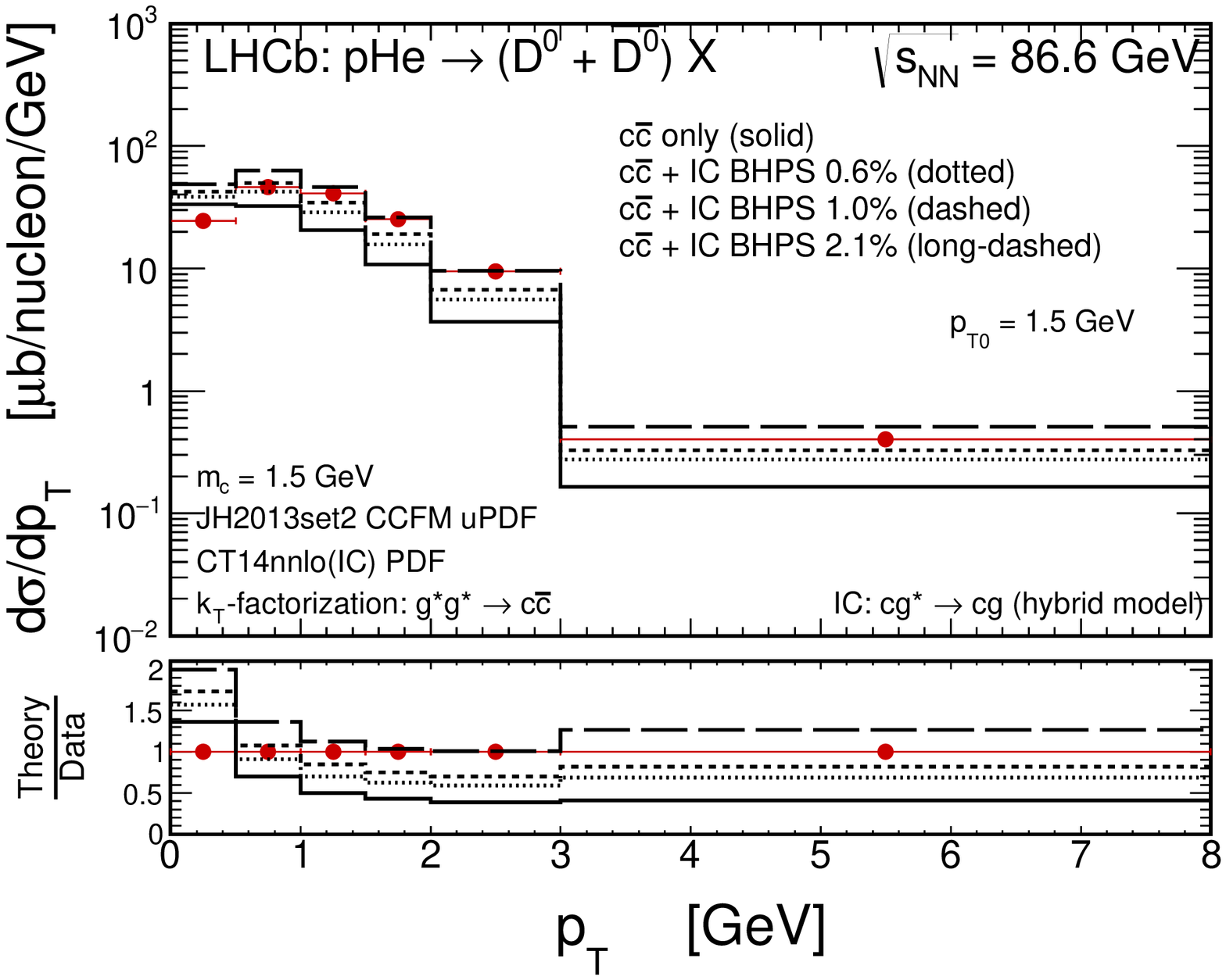}}
\end{minipage}
\begin{minipage}{0.47\textwidth}
  \centerline{\includegraphics[width=1.0\textwidth]{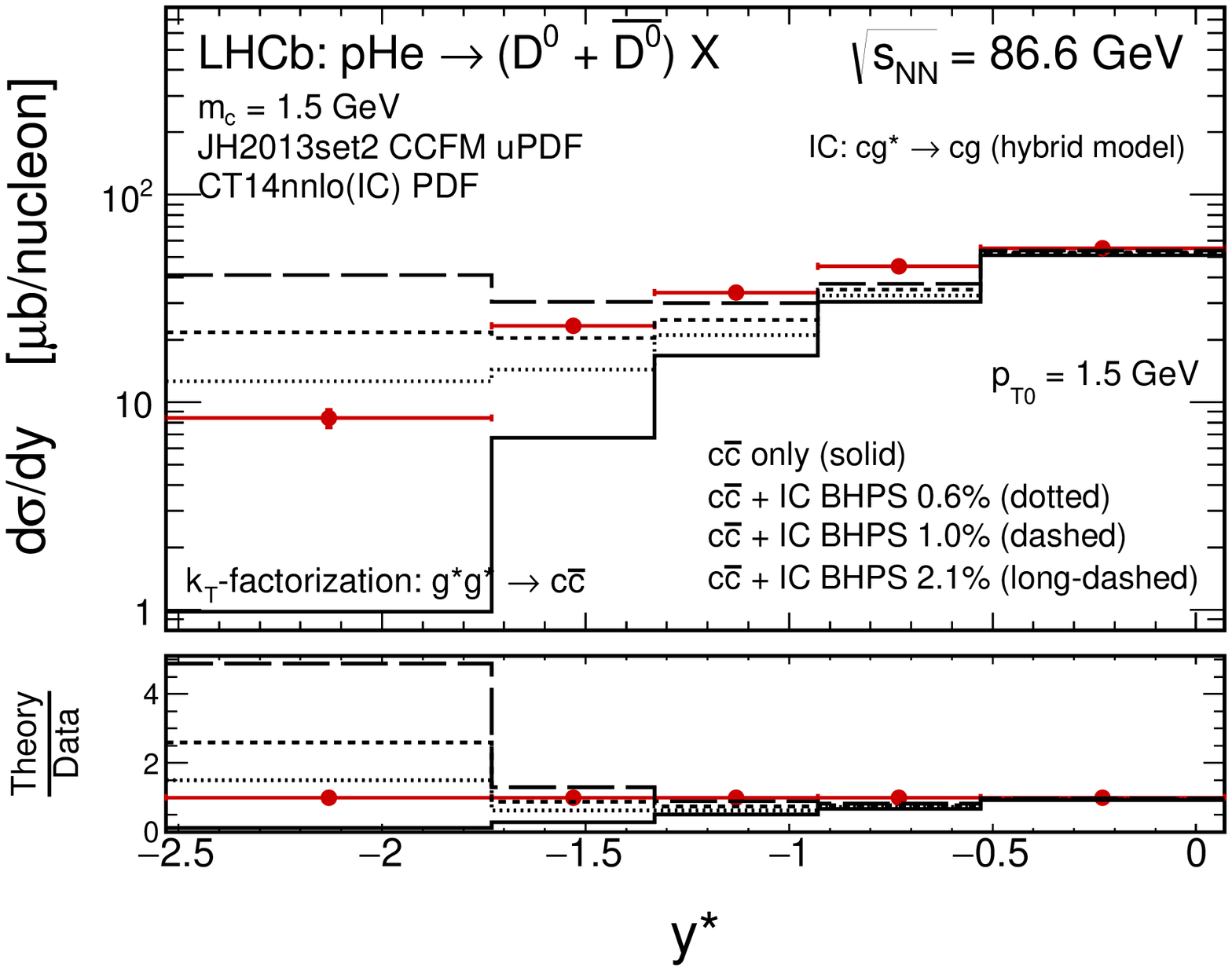}}
\end{minipage}
  \caption{
\small The transverse momentum (left) and rapidity (right) distributions of $D^{0}$ meson (plus $\overline{D^{0}}$ antimeson)
for $p+^4\!\mathrm{He}$ collisions together with the LHCb data \cite{Aaij:2018ogq}. Here results of the standard calculations of the $c\bar c$-pair production obtained with the $k_{T}$-factorization framework and JH2013set2 gluon uPDF are shown without any intrinsic charm component (solid histograms) as well as with different intrinsic charm contributions added (dotted, dashed and long-dashed histograms). Here the IC results correspond to the BHPS model. Other details are specified in the figure.  
}
\label{fig:10}
\end{figure}

\begin{figure}[!h]
\begin{minipage}{0.47\textwidth}
  \centerline{\includegraphics[width=1.0\textwidth]{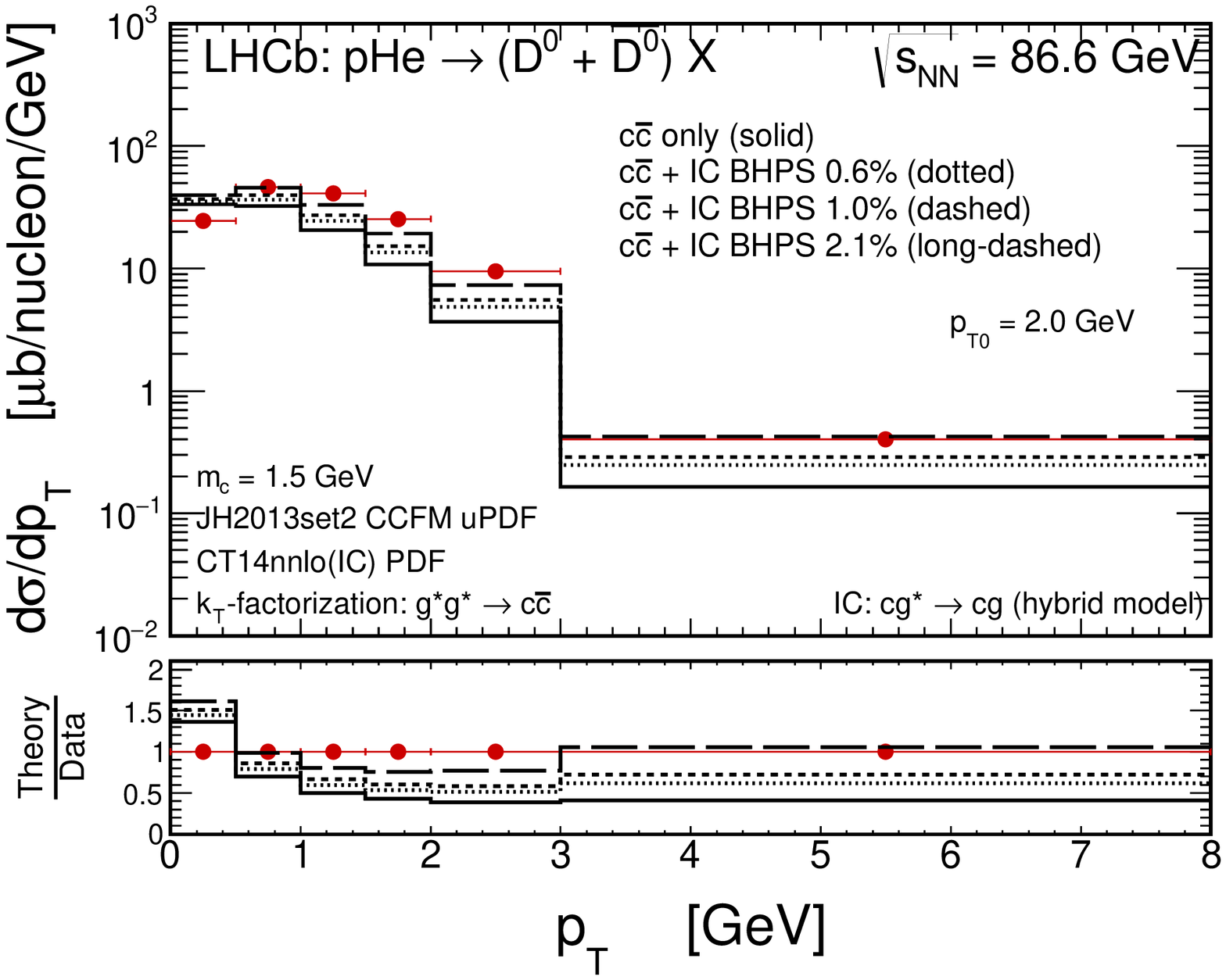}}
\end{minipage}
\begin{minipage}{0.47\textwidth}
  \centerline{\includegraphics[width=1.0\textwidth]{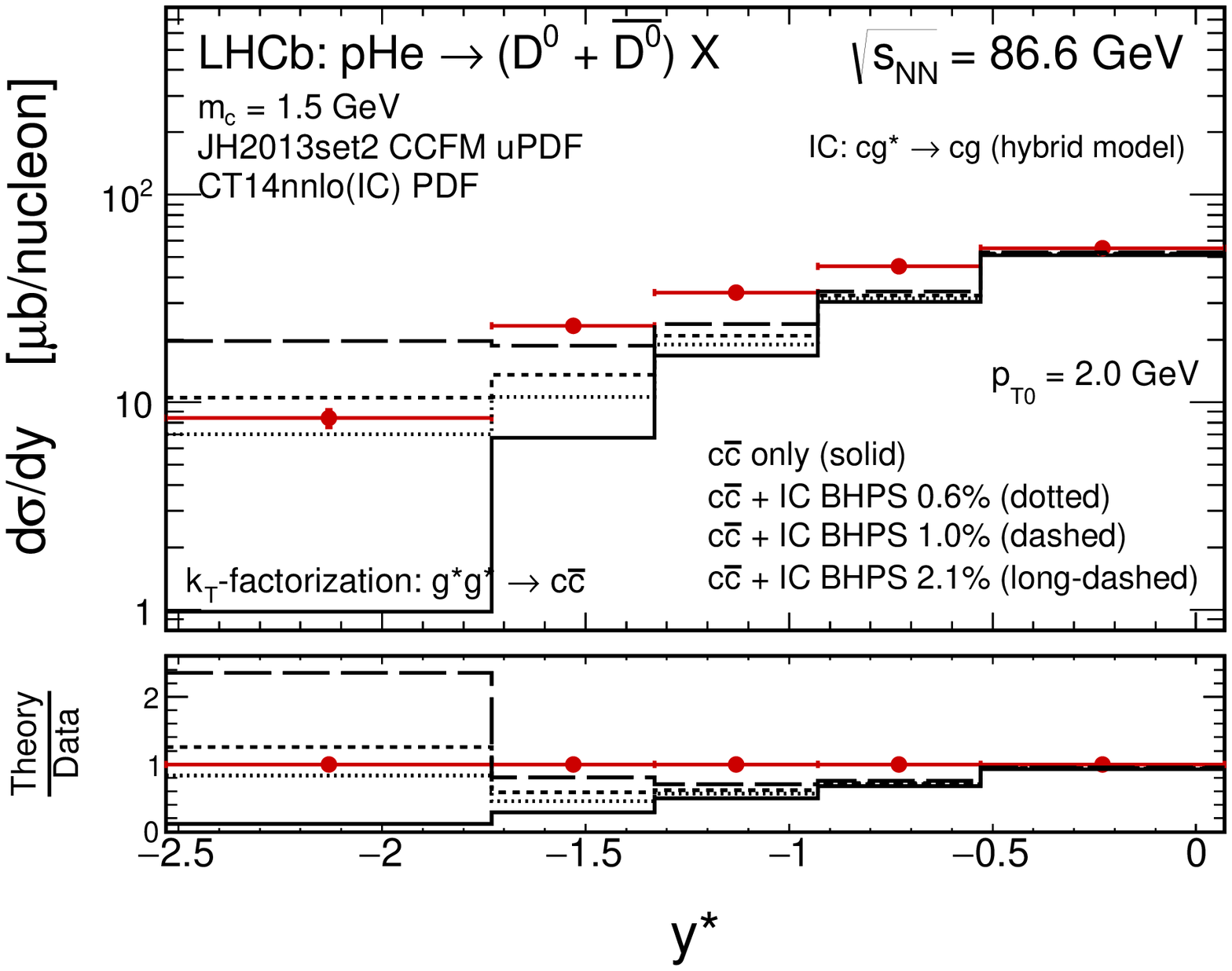}}
\end{minipage}
  \caption{
\small The same as in Fig.~\ref{fig:10} but here the $p_{T0} = 2.0$ GeV was used in the numerical calculations of the intrinsic charm component.  
}
\label{fig:11}
\end{figure}

\begin{figure}[!h]
\begin{minipage}{0.47\textwidth}
  \centerline{\includegraphics[width=1.0\textwidth]{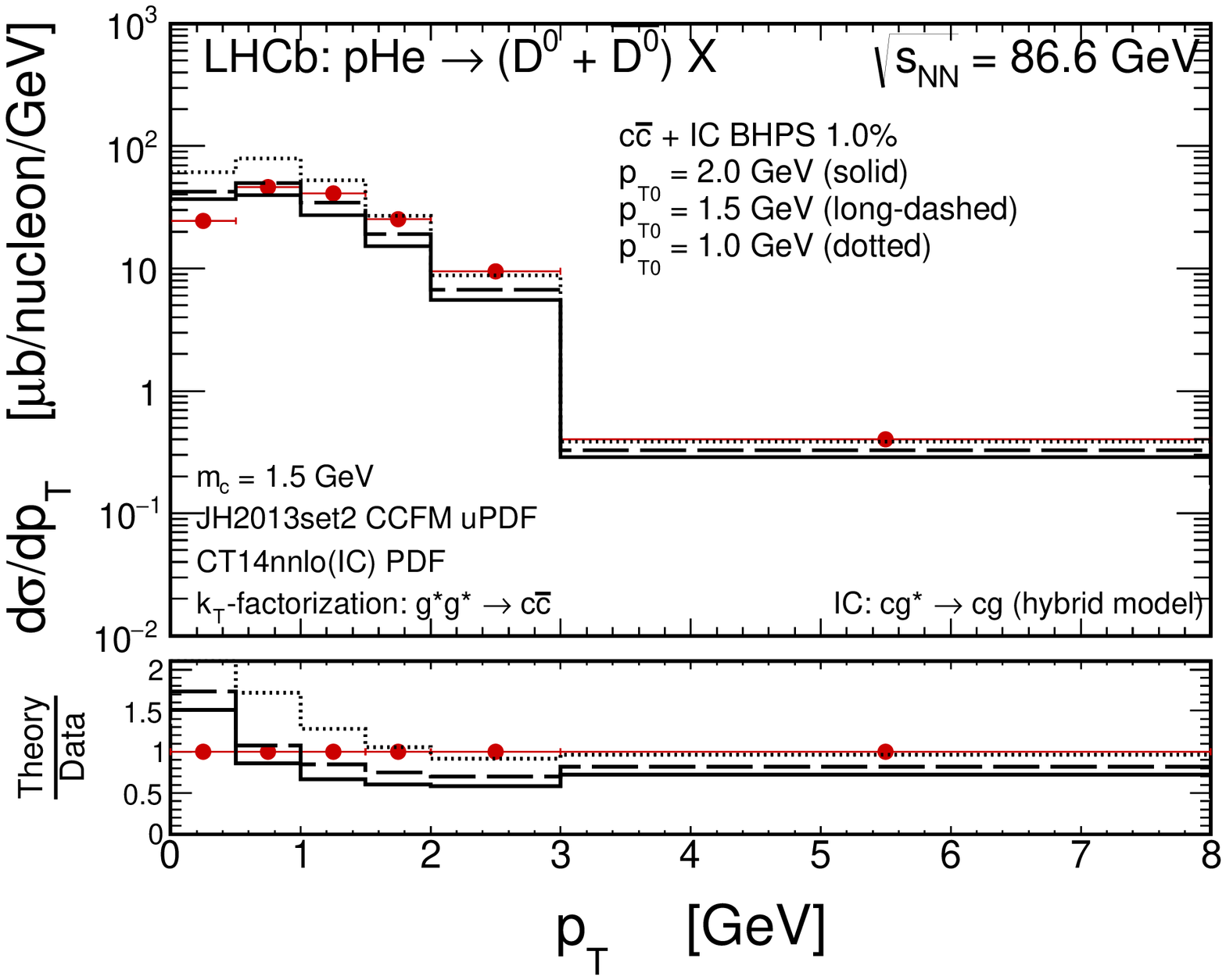}}
\end{minipage}
\begin{minipage}{0.47\textwidth}
  \centerline{\includegraphics[width=1.0\textwidth]{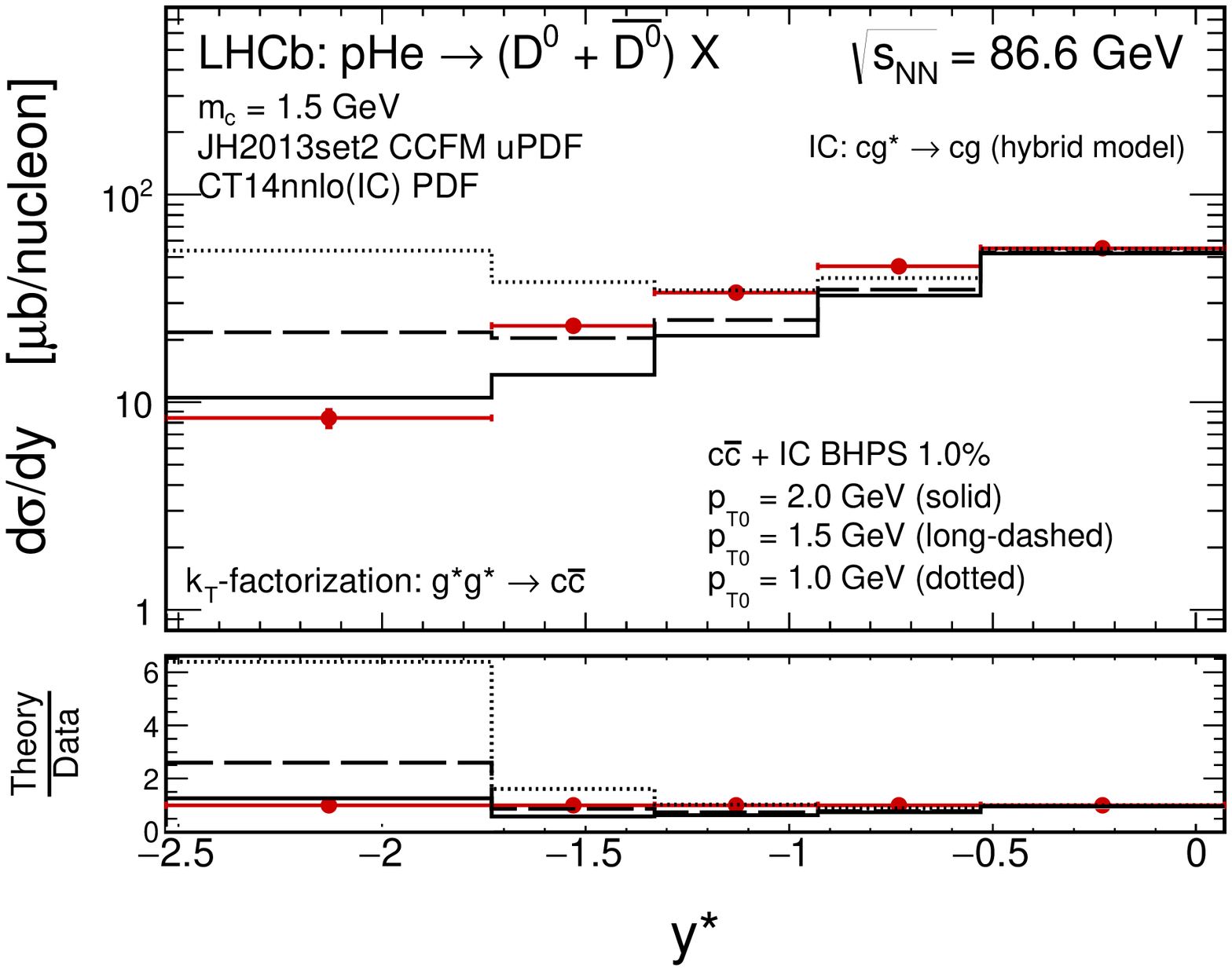}}
\end{minipage}
  \caption{
\small The transverse momentum (left) and rapidity (right) distributions of $D^{0}$ meson (plus $\overline{D^{0}}$ antimeson)
for $p+^4\!\mathrm{He}$ collisions together with the LHCb data \cite{Aaij:2018ogq}. Here we show the sum of the result of the standard calculations of the $c\bar c$-pair production obtained with the $k_{T}$-factorization and JH2013set2 gluon uPDF and the intrinsic charm component for the BHPS model with 1\% probability. Here we show results for three different values of the regularization parameter $p_{T0}$. Other details are specified in the figure. 
}
\label{fig:12}
\end{figure}

As a next step we took into consideration the BHPS model of the
intrinsic charm. In Fig.~\ref{fig:10} again results of the standard
calculations of the $c\bar c$-pair production are shown without any
intrinsic charm component (solid histograms) as well as with different
intrinsic charm contributions added. The dotted, dashed and dash-dotted
histograms correspond to the BHPS model with the $P = 0.6\%$, $P =
1.0\%$, and $P = 2.1\%$, respectively. In contrast to the SEA-like case,
the predictions of the BHPS model contribute significantly to the considered cross section especially at larger meson transverse momentum and in the region of the backward rapidity.
Here the intrinsic charm contribution does not really affect the
standard results at midrapidity and small meson transverse momentum
which is more supported by the LHCb experimental data. A presence of the
intrinsic charm component improves description of the data within the
considered scenario. We see that the $P=0.6\%$ scenario is
favoured by the LHCb fixed-target data. However, as can be seen from 
Fig.~\ref{fig:11}, this conclusion depends on the model parameter 
$p_{T0}$ used in the calculation. For the larger value of 
the regularization parameter, i.e. $p_{T0} = 2.0$ GeV, 
the $P_{IC}=1.0\%$ scenario seems to be more appropriate.
This sensitivity almost disappears for large meson transverse momentum. 
The predicted intrinsic charm contribution for the last bin in $p_{T}$ 
does not really depend on the choice of the $p_{T0}$ what is explicitly
shown in Fig.~\ref{fig:12}. Therefore the last point of the measured
$p_{T}$ spectrum can be used to constrain the $P_{IC}$ probability. 
On the other hand, having the $P_{IC}$ fixed by the data for transverse momentum distribution, the corresponding spectrum in rapidity (especially the very last backward point) could be used to constrain the regularization parameter $p_{T0}$, which is the only free parameter in our model. The rapidity distribution measured by the LHCb experiment strongly supports larger values of the regularization parameter, namely the $p_{T0} = 2.0$ GeV.

As one can see from above, each of the three different values of the intrinsic charm probability visibly modify the $p_{T}$-slope of the differential distribution as well as the backward rapidity region. A presence of the IC components improves quality of the experimental data description. Within the presented theoretical model it is possible to roughly set the upper limit for the intrinsic charm probability to be $\approx 1\%$ which is consistent with the central prediction of the CT14nnloIC PDF global fit.  

\begin{figure}[!h]
\begin{minipage}{0.47\textwidth}
  \centerline{\includegraphics[width=1.0\textwidth]{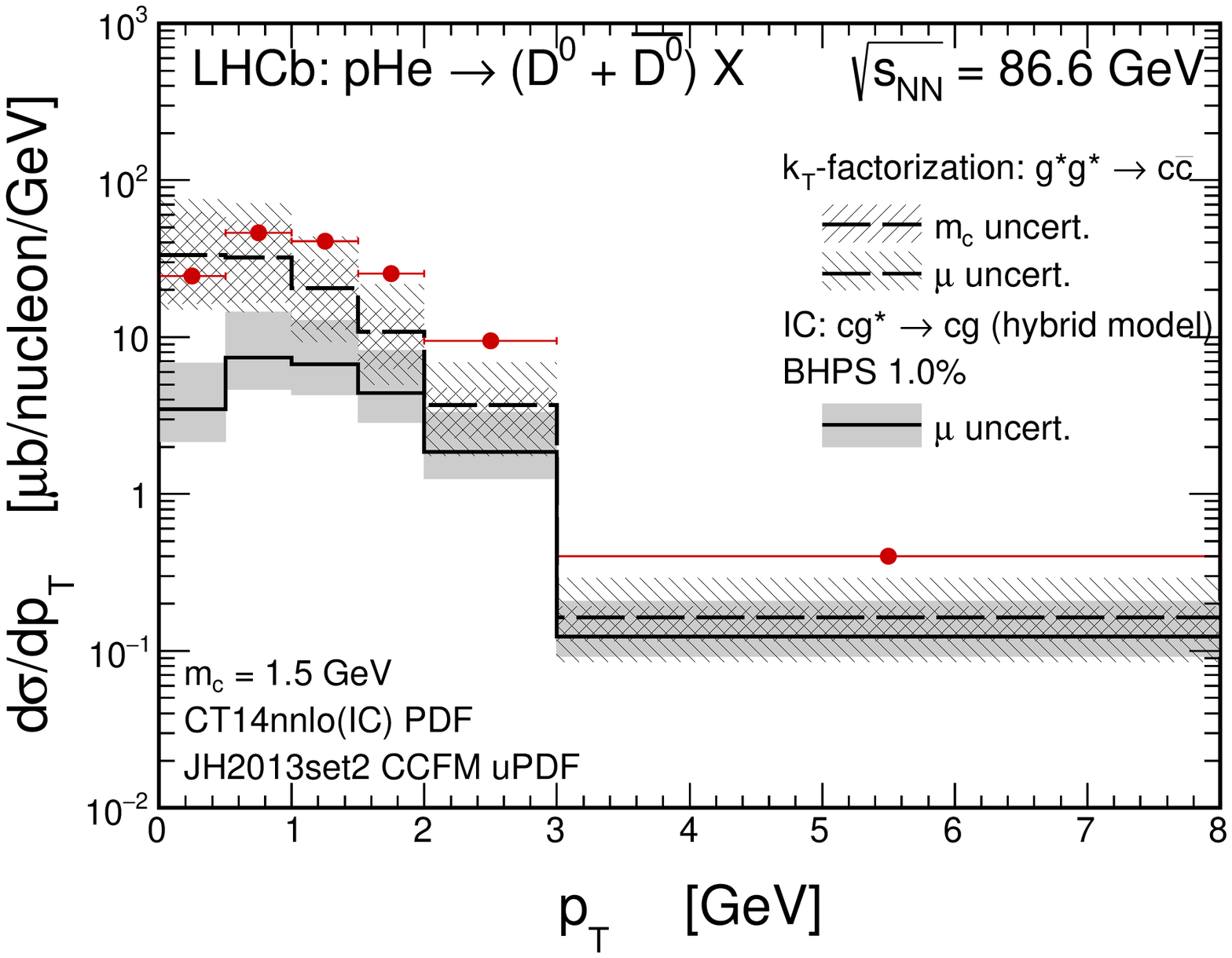}}
\end{minipage}
\begin{minipage}{0.47\textwidth}
  \centerline{\includegraphics[width=1.0\textwidth]{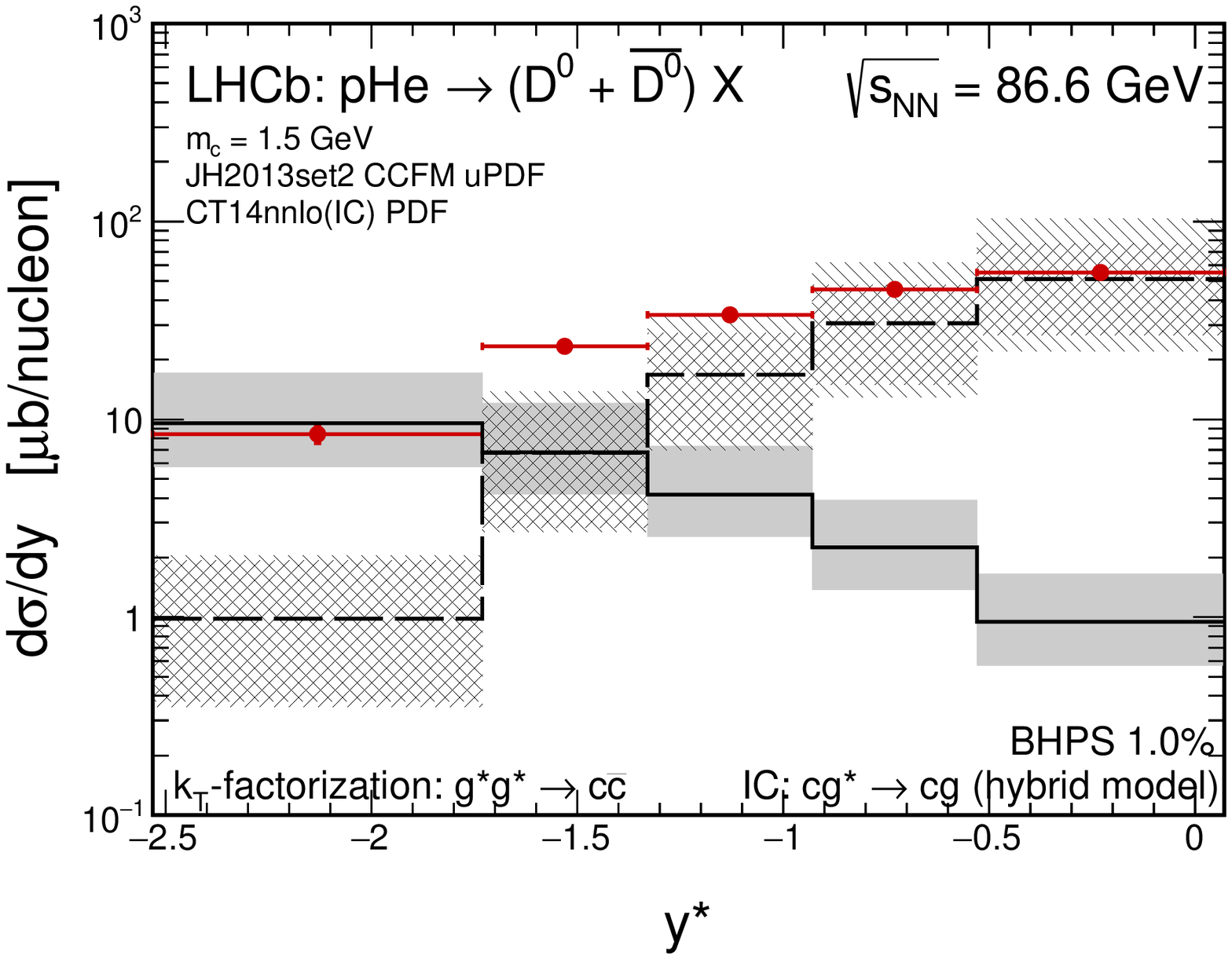}}
\end{minipage}
  \caption{
\small The charm quark mass and the scale uncertainties of our predictions for the standard calculations of the $c\bar c$-pair production obtained with the $k_{T}$-factorization and JH2013set2 gluon uPDF (hatched bands) and the intrinsic charm component for the BHPS model with 1\% probability (shaded bands). Other details are specified in the figure. 
}
\label{fig:uncertainty}
\end{figure}

\begin{figure}[!h]
\begin{minipage}{0.47\textwidth}
  \centerline{\includegraphics[width=1.0\textwidth]{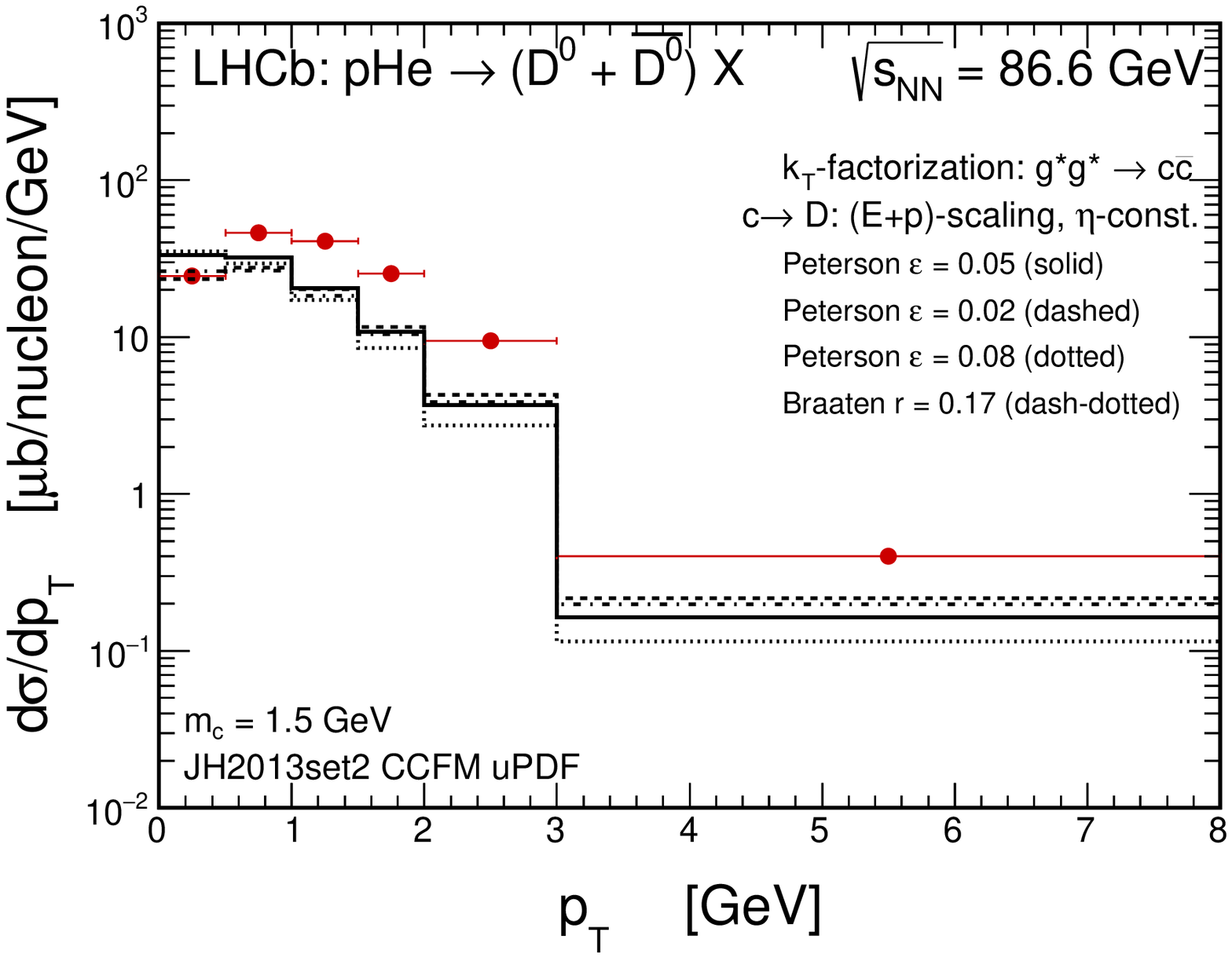}}
\end{minipage}
\begin{minipage}{0.47\textwidth}
  \centerline{\includegraphics[width=1.0\textwidth]{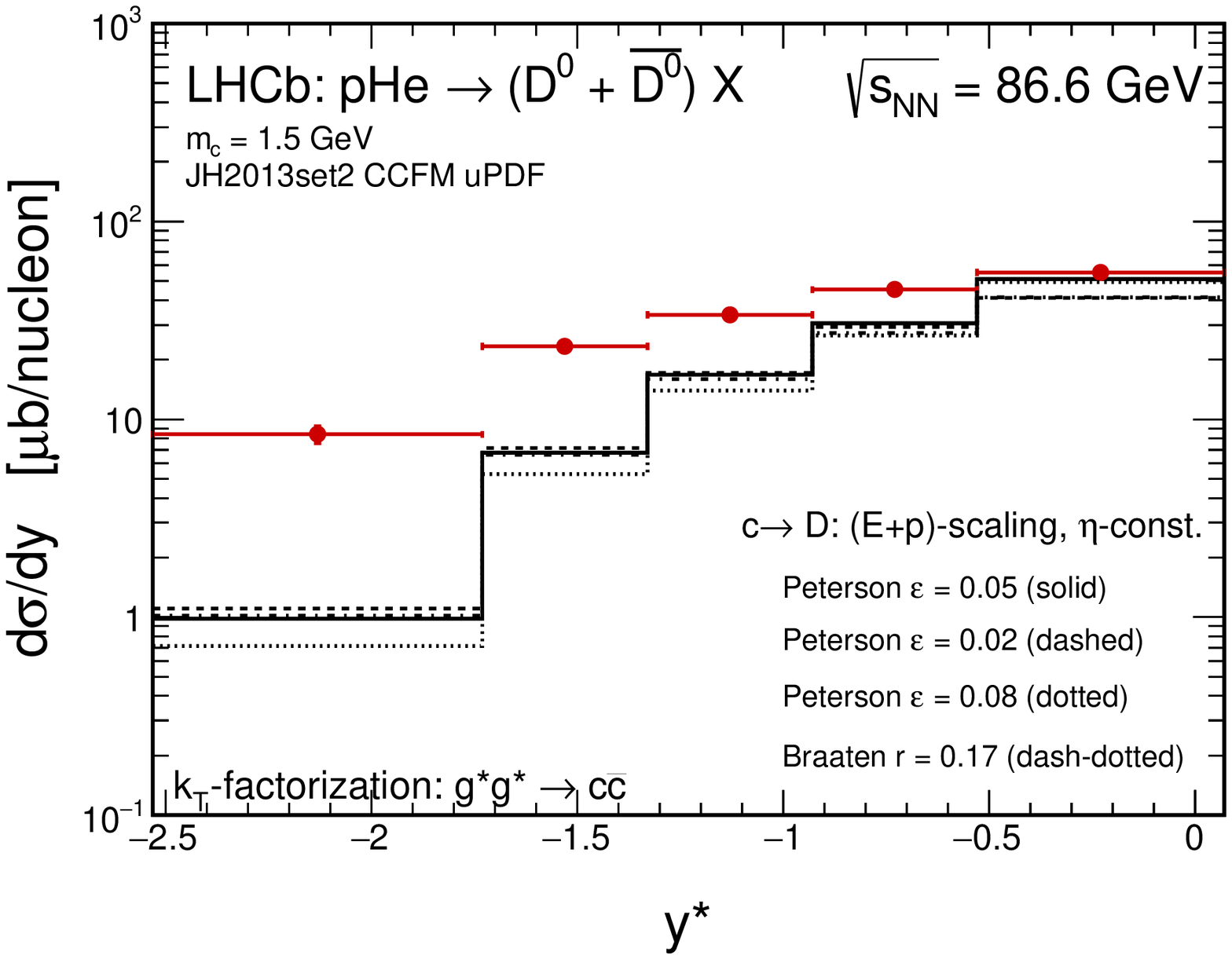}}
\end{minipage}
  \caption{
\small The fragmentation function uncertainties of our predictions for the standard calculations of the $c\bar c$-pair production obtained with the $k_{T}$-factorization and JH2013set2 gluon uPDF. Other details are specified in the figure. 
}
\label{fig:uncertainty-FF}
\end{figure}

\begin{figure}[!h]
\begin{minipage}{0.47\textwidth}
  \centerline{\includegraphics[width=1.0\textwidth]{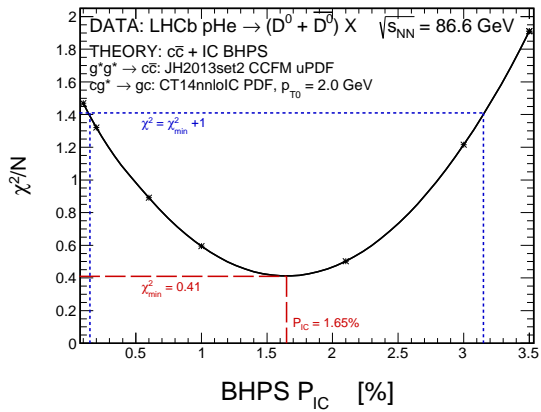}}
\end{minipage}
  \caption{ $\chi^{2}$ as a function of the intrinsic charm probability.
\small 
}
\label{fig:chi2}
\end{figure}

We wish to discuss now uncertainties related to our calculations.
Below we discuss the scale, mass
  (Fig.~\ref{fig:uncertainty}) and the fragmentation function
  (Fig.~\ref{fig:uncertainty-FF}) uncertainties 
as well as the $\chi^{2}$ analysis (Fig.~\ref{fig:chi2}) used in order
to set limits on the IC probability.

In Fig.~\ref{fig:uncertainty} we show uncertainties related to quark
mass and the choice of the scale. 
The uncertainties are not small, however the shapes of distributions,
particularly that for $d \sigma / dy$, are quite different than the
experimental ones and the IC contribution visibly improves description of the rapidity distribution.

In Fig.~\ref{fig:uncertainty-FF} we show the uncertainties due to
the choice of the fragmentation functions. Such uncertainties are much
smaller than those due to scale or charm quark mass and can be safely neglected.

In Fig.~\ref{fig:chi2} we show $\chi^2$ as a function of the intrinsic charm
probability $P_{IC}$. Here we follow the method presented in Refs.~\cite{Bednyakov:2017vck,Brodsky:2020zdq}
and define $\chi^2$ as
\begin{equation}
\chi^2(P_{IC}) = \sum_{i=1}^{n} \frac{[y_{i} - f(P_{IC})_{i}]^2}{\Delta\sigma_{i}^2},
\label{chi2}
\end{equation}
where $y_{i}$ is the measurement, $f(P_{IC})_{i}$ is the theoretical result for a given $P_{IC}$ and $\Delta\sigma_{i}$ is is the sum in quadrature of the uncertainties coming from the measurement and the uncertainties coming from the theoretical calculations.
We observe a minimum at $P_{IC} \sim$ 1.65 \% but the uncertainty is
really large. A large range of the IC probability is still allowed.
In parallel we made a study of extracting the IC component
from the IceCube neutrino data. There only upper limit $P_{IC} \lesssim$ 2.0\%
could be extracted \cite{Goncalves:2021yvw}. This seems consistent with the present result.
Similar result was obtained in Ref.~\cite{Bednyakov:2017vck}.

\section{Conclusions}

In the present paper we have discussed in detail how the relatively
recent data on $D^0/{\bar D}^0$ production in $p\!+\!^{4}\!He$
fixed-target LHCb experiment put constraints on intrinsic charm component
in the nucleon. We have considered two models of $c \bar c$ symmetric
intrinsic charm: the well known BHPS model as well as a SEA-like model.
In our approach the knocked out $c$-quark or ${\bar c}$-antiquark
fragment producing $D^0$ or ${\bar D}^0$ mesons.
This part of the reaction is described by a phenomenological
Peterson fragmentation function correctly using kinematics.

As a "background" contribution we have used the traditional gluon-gluon
and quark-antiquark annihilation mechanism. The respective contributions
were calculated within the $k_T$-factorization approach. The gluon-gluon
fusion was calculated using so-called CCFM unintegrated gluon
distribution adjusted to experimental $F_2$ structure function.
The quark-antiquark annihilation contributions turned out negligible
compared to the sum of gluon-gluon and IC contributions.

Within the scenario presented here the traditional components seem 
insufficient to describe the LHCb data, especially when discussing the 
very backward rapidity region and large meson transverse momenta.
Here we follow a different treatment of the fragmentation process 
with respect to the predictions presented by one of us in
Ref.~\cite{Maciula:2020cfy}. A corresponding consequences related to 
the modelling o hadronization effects at lower energies are 
carefully discussed.

We have studied uncertainties of our calculation related to
scale and charm quark mass as well as $c \to D$ fragmentation functions. 
The first two are large. However, the shape of rapidity distribution of 
the traditional calculation is rather different than the experimental one.
The uncertainty due to fragmentation functions seems rather small.
The BHPS IC improves the description of the data. We have also tried to
set limits on the probability of the IC component. We have observed
a minimum of $\chi^2$ at $P_{IC} \approx$ 1.65 \%.
There is, however, a big uncertainty and the probability 0.2-3.15 \%
within $1\sigma$ confidence level. For comparison the upper limit obtained from the analysis
of the neutrino IceCube data \cite{Goncalves:2021yvw} is $P_{IC} \lesssim 2.0 \%$ , which is
consistent with the present analysis. Similarly, in Refs.~\cite{Bednyakov:2017vck,Brodsky:2020zdq} the authors
found $P_{IC} \lesssim 1.93 \%$ using tha ATLAS data on the associated production of prompt photons and charm-quark jets.



\vskip+5mm
{\bf Acknowledgments}\\
We are indebted to Aleksander Kusina for discussion of nuclear effects.
This work has been partially supported by the Mainz Institute for 
Theoretical Physics (MITP) of the Cluster of Excellence PRISMA+ 
(Project ID 39083149).
This study was also partially supported by the Polish National Science Center
grant UMO-2018/31/B/ST2/03537
and by the Center for Innovation and
Transfer of Natural Sciences and Engineering Knowledge in Rzesz{\'o}w.


\end{document}